\documentstyle[12pt]{article}
\newcommand{\mysection}{\setcounter{equation}{0}\section}

\begin{document}
\begin {flushright}
ITP-SB-95-16
\end {flushright}
\vspace{3mm}
\begin{center}
{\Large \bf Bottom quark production cross section at fixed-target
$pp$ experiments}
\end{center}
\vspace{2mm}
\begin{center}
N. Kidonakis and J. Smith  \\
\vspace{2mm}
{\it Institute for Theoretical Physics,
State University of New York at Stony Brook,
Stony Brook, NY 11794-3840, USA} \\
\vspace{2mm}
June 5, 1995
\end{center}

\begin{abstract}
The cross section for bottom quark production at fixed-target energies is
calculated for a wide range of beam momenta.
A detailed analysis is given for the HERA-B experiment.
We consider both the
order $\alpha_s^3$ cross section and the resummation of
soft gluon corrections in all orders of QCD perturbation theory.
The inclusive transverse momentum and rapidity distributions,
including resummation, for
bottom quark production at HERA-B are also presented.
\end{abstract}

\pagebreak
%------------------This is Section 1---------------------------------
\mysection{Introduction}
%----------------------------------------------------------
The calculation of production cross sections for heavy particles
in QCD is made by invoking the factorization theorem \cite{css} and
expanding the contributions to the amplitude in powers of the coupling
constant $\alpha_s(\mu^2)$. Recent investigations have
shown that near threshold there can be large logarithms in the perturbation
expansion which have to be resummed to make more reliable theoretical
predictions. The application of these ideas to fixed-target
Drell-Yan production has been the subject of many papers over
the past few years \cite{gs}.
The same ideas on resummation were applied to the calculation of
the top-quark cross section at the Fermilab Tevatron in \cite{LSN} and
\cite{lsn2}.
What is relevant in these reactions
is the existence of a class of logarithms of the type
$(\ln(1-z))^{i}/(1-z)$, where $i$ is the order of the
perturbation expansion, and where one must integrate over the variable
$z$ up to a limit $z=1$. These
terms are not actually singular at $z = 1$ due to the presence
of terms in $\delta(1-z)$. However the remainder
can be quite large. In general one writes such terms
as ``plus'' distributions, which are then convoluted with
regular test functions (the parton densities).

In this paper we examine the production of $b$-quarks
in a situation where the presence of these large logarithms is
of importance, namely in a fixed-target experiment to be performed
in the HERA ring at DESY. This actual experiment has the name HERA-B
\cite{herab1,herab2}
and involves colliding the circulating proton beam against a stationary
copper wire in the beam pipe. The nominal beam energy of the protons
is 820 GeV, so that the square root of the center-of-mass (c.m.) energy
is $\sqrt{S} = 39.2 $ GeV. Taking the $b$-quark mass as
$m_b = 4.75\: {\rm GeV}/c^2$ then the ratio of $m_b/\sqrt{S} \approx 1/8$.
If we choose the renormalization scale in the running coupling constant
as $m_b$ then $\alpha_s(m_b^2) \approx 0.2$ so
$\alpha_s(m_b^2) \ln(\sqrt{S}/m_b) \approx  0.4$. This number is
small enough that we expect a reasonably convergent perturbation series.

In perturbation theory with a hard scale we can use the
standard expression for the order-by order cross section
in QCD, namely
\begin{equation}
\sigma(S,m^2) = \int_{4m^2}^1 dx_1 \int_{\frac{4m^2}{Sx_1}}^1 dx_2 \sum_{ij}
f_i(x_1,\mu^2) f_j(x_2,\mu^2) \sigma_{ij}(s = x_1 x_2 S, m^2,\mu^2),
\end{equation}
where the $f_i(x,\mu^2)$ are the parton densities at the factorization
scale $\mu^2$ and the $\sigma_{ij}$ are the partonic cross sections.
The numerical results for the hadronic cross sections
depend on the choice of the
parton densities, which involves the mass factorization
scale $\mu^2$; the choice of the running coupling constant, which
involves the renormalization scale (also normally chosen to be $\mu^2$);
and the choice for the actual mass of the $b$-quark.
In lowest order or Born approximation the
actual numbers for the cross section show a large sensitivity to
these parameters.  In section 3 we will show plots of
the production cross section in
leading order (LO), i.e. $O(\alpha_s^2)$,
and next-to-leading order (NLO), i.e. $O(\alpha_s^3)$.
The NLO results follow from the work of the
two groups \cite{nde1} and \cite{{betal},{bnmss}}.
However even including the NLO
corrections does not completely fix the cross section. The sensitivity
to our lack of knowledge of even higher terms in the QCD expansion is usually
demonstrated by varying the scale choice up and down by factors
of two. In general it is impossible to make more precise
predictions given the absence of a calculation in next-to-next-to-leading
order (NNLO). However in specific kinematical regions we can do
so.

The threshold region is one of these regions.
In this region one finds that there are large logarithms which arise
from an imperfect cancellation
of the soft-plus-virtual (S+V) terms in the perturbation expansion.
These logarithms are exactly of the same type mentioned above.
We will see in section 3 that the gluon-gluon channel
is the dominant channel for the production
of $b$-quarks near threshold in a fixed-target $pp$ experiment.
This is not the case for the production of the top quark at the
Fermilab Tevatron, which is a proton-antiproton collider, and where
the dominant channel is the quark-antiquark one. That was fortunate
as the exponentiation of the soft-plus-virtual terms in \cite{LSN}
is on a much more solid footing in the $q \bar q$ channel, due to all
the past work which has been done on the Drell-Yan reaction \cite{gs}.
Since the gluon-gluon channel is now the most
important one we are forced to reexamine all ``large''
corrections near threshold, including both Coulomb-like
and large constant terms. We will do that
in section 2 where we will present all the relevant formulae at
the partonic level and will
also discuss the exponentiation of these terms.
In addition we will present subleading S+V terms and discuss their
contribution to the total S+V cross section.
Section 3 contains the analysis of the hadron-hadron cross section
which is relevant for the HERA-B experiment as well as for fixed-target
$pp$ experiments in general. We give
results in LO, in NLO and after resummation.
Finally in Section 4 we give our conclusions and discuss where more
work should be done in the future.

%------------------This is Section 2---------------------------------
\mysection{Results for parton-parton reactions}
%------------------------------------------------------------------
The partonic processes that we examine are
\begin{equation}
i(k_1) + j(k_2) \rightarrow Q(p_1) + \bar Q(p_2),
\end{equation}
where $i,j = g,g$ or $i,j= q, \bar q$ and $Q, \bar Q$ are heavy quarks
$(c,b,t)$.
The square of the parton-parton c.m. energy is
$s=(k_1+k_2)^2$.

We begin with an analysis of heavy quark production in the $q \bar q$ channel.
The Born cross section in this channel is given by
\begin{equation}
\sigma_{q \bar q}^{(0)}(s,m^2)=\frac{2 \pi}{3} \alpha_s^2(\mu^2)
 K_{q \bar q} N C_F \frac{1}{s} \beta \left(1+\frac{2m^2}{s}\right),
\end{equation}
where $C_F=(N^2-1)/(2N)$ is the Casimir invariant for the fundamental
representation of $SU(N)$,
$K_{q \bar q}=N^{-2}$ is a color average factor,
$m$ is the heavy quark mass, $\mu$ denotes the
renormalization scale, and $\beta=\sqrt{1-4m^2/s}$.
Also $N=3$ for the $SU(3)$ color group in QCD.
The threshold behavior ($s \rightarrow 4m^2$) of this expression is given by
\begin{equation}
\sigma_{q \bar q ,\, \rm thres}^{(0)}(s,m^2)=\pi\alpha_s^2(\mu^2)
K_{q \bar q} N C_F \frac{1}{s} \beta.
\end{equation}
Complete analytic results are not available for the NLO cross section
as some integrals are too complicated to do by hand.
However in \cite{bnmss} analytic results are given for
the soft-plus-virtual contributions to the cross section, and for the
approximation to the cross section near threshold.
Simple formulae which yield reasonable approximations to the exact
$O(\alpha_s^3)$ results have been constructed in \cite{MSSN}. From these
results one can derive that the Coulomb terms to first order in the
$q \bar q$ channel are given by
\begin{equation}
\sigma^{(\pi^2)}_{q \bar q}(s,m^2)=\sigma^{(0)}_{q \bar q}(s,m^2)
\frac{\pi \alpha_s(\mu^2)}{2\beta}\left(C_F-\frac{C_A}{2}\right)
\end{equation}
in the $\overline {\rm MS}$ scheme,
where $C_A=N$ is the Casimir invariant for the adjoint representation
of $SU(N)$.
These terms are distinguished by their typical
$\beta^{-1}$ behaviour near threshold which, after multiplication by the Born
cross section, yield finite cross sections at threshold in NLO.
We note that $C_F-C_A/2=-1/6$ is a negative quantity for $SU(3)$
and that the first-order
Coulomb correction is negative (the interaction is repulsive).
{}From the original work of Schwinger \cite{Schwinger}
we know that the Coulomb terms
exponentiate. Therefore we resum them by writing
\begin{equation}
\sigma^{(\pi^2),\, \rm res}_{q \bar q}(s,m^2)= \sigma^{(0)}_{q \bar q}(s,m^2)
\exp\left[\frac{\pi \alpha_s(\mu^2)}{2\beta}
\left(C_F-\frac{C_A}{2}\right)\right].
\end{equation}
Since the exponent is negative, the exponentiation of the Coulomb terms
actually supresses the cross section, and the resummed result
goes to zero at threshold.
In a previous treatment of threshold effects
\cite{FKS} the exponentiation was done in a different way,
namely by writing
\begin{equation}
\sigma^{(\pi^2),\, \rm res}_{q \bar q}(s,m^2)= \sigma^{(0)}_{q \bar q}(s,m^2)
\frac{X_{(8)}}{\exp X_{(8)}-1},
\end{equation}
where
\begin{equation}
X_{(8)}=-\frac{\pi \alpha_s(\mu^2)}{\beta}
\left(C_F-\frac{C_A}{2}\right)=\frac{1}{6}\frac{\pi\alpha_s}{\beta}.
\end{equation}
The subscript in $X_{(8)}$ indicates that in the $q\bar q$ channel,
where the process has to go via an intermediate $s$-channel gluon,
the heavy quark pairs
are produced exclusively in the color octet state.
We have checked that
the difference between the two methods of resummation is numerically
negligible.

In the DIS scheme in addition to the Coulomb terms we also have
a large constant contribution so that the
first order result near threshold is
\begin{equation}
\sigma^{(\pi^2)}_{q \bar q}(s,m^2)=\sigma^{(0)}_{q \bar q}(s,m^2)
\left[\frac{\pi\alpha_s(\mu^2)}{2\beta}\left(C_F-\frac{C_A}{2}\right)
+\frac{\alpha_s(\mu^2)}{\pi}C_F\left(\frac{9}{2}+\frac{\pi^2}{3}\right)\right].
\end{equation}
We know that large constants exponentiate in the Drell-Yan reaction
\cite{gs} so we assume that the same holds here and write the result in
the DIS scheme as
\begin{eqnarray}
\sigma^{(\pi^2),\, \rm res}_{q \bar q}(s,m^2)&=& \sigma^{(0)}_{q \bar q}(s,m^2)
\exp\left[\frac{\pi \alpha_s(\mu^2)}{2\beta}\left(C_F-\frac{C_A}{2}\right)
\right]
\nonumber \\ &&
\times \exp\left[\frac{\alpha_s(\mu^2)}{\pi}C_F\left(\frac{9}{2}
+\frac{\pi^2}{3}\right)\right].
\end{eqnarray}
This expression also goes to zero at threshold.
We have included the constant terms to see their effect at
larger values of $\beta$.

Since the total parton-parton cross sections only depend on the variables
$s$ and $m^2$ they can be expressed in terms of scaling functions
as follows
\begin{eqnarray}
\sigma_{ij}(s,m^2)&=&\sum_{k=0}^{\infty} \sigma_{ij}^{(k)}(s,m^2)
\nonumber \\ & =&
\frac{\alpha_s^2(\mu^2)}{m^2} \sum_{k=0}^{\infty}
(4\pi\alpha_s(\mu^2))^k \sum_{l=0}^{k} f_{ij}^{(k,l)}(\eta)
\ln^l\frac{\mu ^2}{m^2},
\end{eqnarray}
where we denote by $\sigma^{(k)}$ the $O(\alpha_s^{k+2})$ contribution
to the cross section.
The scaling functions $f_{ij}^{(k,l)}(\eta)$ depend on the scaling variable
$\eta=s/4m^2-1=s\beta^2/4m^2$.

In fig. 1 we plot $f_{q \bar q}^{(k,0)}(\eta)$ for $k=0,1$ for the
exact and threshold  expressions (from \cite{bnmss}) in the
$\overline {\rm MS}$ scheme. We also plot
$f_{q \bar q}^{(\pi^2),\, \rm res}(\eta,\alpha_s)$ which we define by
\begin{equation}
\sigma_{q \bar q}^{(\pi^2),\,\rm res}(s, m^2)=\frac{\alpha_s^2(\mu^2)}{m^2}
f_{q \bar q}^{(\pi^2),\, \rm res}(\eta,\alpha_s).
\end{equation}
We see that the threshold Born approximation is excellent for small
$\eta$ and reasonable for the entire range of $\eta$ shown.
As expected the resummed Coulomb terms suppress the Born result
throughout the entire range of $\eta$ and go to zero at threshold.
We also note that the theshold first-order approximation is good
only very near to threshold.
In fig. 2 we plot the corresponding functions for the DIS scheme.
Here the first-order corrections are smaller than in the
$\overline {\rm MS}$ scheme. Again the threshold first-order approximation
is good only very close to threshold. The resummed result differs greatly
from that in the $\overline {\rm MS}$ scheme. Here the additional
exponentiation of large
constants cancels the negative contribution of the Coulomb terms and
produces a large enhancement of the Born term in the region
$0.1<\eta<1$ which, as we will show in the next section, is the most
important region kinematically for the production of $b$-quarks at HERA-B.
For small $\eta$, however, the total resummed result is dominated
by the pure Coulomb terms and thus suppresses the Born term.
Finally, we also show the resummed result where the only constant
that we exponentiate is the $\pi^2/3$ term in (2.9). In this case the
enhancement of the Born term in our region of interest is much smaller.

The analysis of the contributions to the gluon-gluon channel in NLO
is much more complicated. First of all there are three Born diagrams
each with a different color structure.  Therefore only few terms
near threshold are proportional to the Born cross section.
The exact Born term in the $gg$ channel is
\begin{eqnarray}
\sigma_{gg}^{(0)}(s,m^2)&=&4\pi\alpha_s^2(\mu^2) K_{gg}N C_F \frac{1}{s}
\left\{C_F\left[-\left(1+\frac{4m^2}{s}\right)\beta\right.\right.
  \nonumber \\ && \quad \quad \quad \quad
+\left. \left(1+\frac{4m^2}{s}-\frac{8m^4}{s^2}\right)
 \ln\frac{1+\beta}{1-\beta}\right]
\nonumber \\ &&
+\left.C_A\left[-\left(\frac{1}{3}+\frac{5}{3}\frac{m^2}{s}\right)\beta
+\frac{4m^4}{s^2}\ln\frac{1+\beta}{1-\beta}\right]\right\},
\end{eqnarray}
where $K_{gg}=(N^2-1)^{-2}$ is a color average factor.
The threshold behavior ($s \rightarrow 4m^2$) of this expression is given by
\begin{equation}
\sigma_{gg,\, \rm thres}^{(0)}(s,m^2)= \pi \alpha_s^2(\mu^2) K_{gg}
\frac{1}{s} N C_F [4 C_F-C_A] \beta.
\end{equation}
Again, the complete NLO expression for the cross section in the $gg$ channel
is unavailable but analytic results are given for the S+V terms in
\cite{betal}. These were used in \cite{MSSN} to analyze the magnitude of the
cross section near threshold.
{}From the approximate expressions given in \cite{MSSN}
one can extract the $\pi^2$ terms to first order in the $gg$ channel.
These are
\begin{eqnarray}
\sigma_{gg}^{(\pi^2)}(s,m^2)&=&\alpha_s^3(\mu^2) N C_K K_{gg} \frac{\pi^2}{s}
\left[\frac{5}{8}+\frac{1}{24}\beta^2+16\frac{m^6}{s^3} \right.
\nonumber \\ && \quad \quad \quad
+\left.\left(32\frac{m^8}{s^4}-10\frac{m^4}{s^2}\right)\frac{1}{\beta}
\ln\frac{1+\beta}{1-\beta}\right] \nonumber \\ &&
+\alpha_s^3 C_{\rm QED} K_{gg} \frac{\pi^2}{s}
\left[-\frac{1}{4}-16\frac{m^6}{s^3} \right. \nonumber \\ && \quad \quad \quad
\left.+\left(-32\frac{m^8}{s^4}+8\frac{m^4}{s^2}\right)\frac{1}{\beta}
\ln\frac{1+\beta}{1-\beta}\right],
\end{eqnarray}
where $C_K=(N^2-1)/N=2N C_F C_A-4N C_F^2$, and $C_{\rm QED}=(N^4-1)/N^2
=-4C_F^2+4C_A C_F$.
These are not proportional to the Born term so that it is
not clear how to resum them.
The threshold behavior of (2.14) is given by
\begin{equation}
\sigma_{gg,\, \rm thres}^{(\pi^2)}(s,m^2)=\alpha_s^3(\mu^2) K_{gg}
\frac{\pi^2}{4}\frac{1}{s}\left[\frac{-N C_K}{2}+C_{\rm QED}\right],
\end{equation}
which is proportional to the
threshold Born term.
Therefore the threshold approximation for the $\pi^2$ terms
in NLO can be written as
\begin{equation}
\sigma_{gg,\, \rm thres}^{(0)+(\pi^2)}(s,m^2)=
\sigma_{gg,\, \rm thres}^{(0)}(s,m^2) \left[1+\frac{\pi\alpha_s(\mu^2)}
{4\beta}\left(\frac{-NC_K/2+C_{\rm QED}}{(4C_F-C_A)N C_F}\right)\right],
\end{equation}
or, writing the color factors in terms of $N$, as
\begin{equation}
\sigma_{gg,\, \rm thres}^{(0)+(\pi^2)}(s,m^2)=
\sigma_{gg,\, \rm thres}^{(0)}(s,m^2) \left[1+\frac{\pi\alpha_s(\mu^2)}
{4\beta}\frac{N^2+2}{N(N^2-2)}\right].
\end{equation}
The correction is positive indicating a Coulomb attraction in this channel.
Then we proceed to resum these terms by the Schwinger method \cite{Schwinger}
\begin{equation}
\sigma_{gg,\, \rm thres}^{(\pi^2),\, \rm res}(s,m^2)=
\sigma_{gg,\, \rm thres}^{(0)}(s,m^2)
\frac{X}{1-\exp(-X)} \, ,
\end{equation}
where
\begin{eqnarray}
X&=& \frac{\pi\alpha_s(\mu^2)}{2 \beta }
\frac{N^2+2}{N(N^2-2)}
\nonumber \\ &=&
\frac{11}{42}\frac{\pi\alpha_s}{\beta} \:\:{\rm for} \:\:SU(3).
\end{eqnarray}
The reason that we have exponentiated in this way is that $X$ is positive
and tends to infinity when $\beta\rightarrow 0$.
We expect that the Coulomb terms are only important very close
to threshold.
In \cite{FKS} the Coulomb singlet (attractive interaction)
and octet (repulsive interaction) contributions to the $gg$ channel
were exponentiated separately giving
\begin{equation}
\sigma_{gg}^{(s)}=\frac{2}{7}\sigma_{gg}^{(0)}
\frac{X_{(s)}}{1-\exp(-X_{(s)})},
\quad  X_{(s)}=\frac{4}{3}\frac{\pi\alpha_s}{\beta},
\end{equation}
and
\begin{equation}
\sigma_{gg}^{(8)}
=\frac{5}{7}\sigma_{gg}^{(0)}
\frac{X_{(8)}}{\exp(X_{(8)})-1},
\quad X_{(8)}=\frac{1}{6}\frac{\pi\alpha_s}{\beta},
\end{equation}
respectively. Again we have checked that the difference
between the two methods of exponentiation is not significant numerically.

In fig.3 we plot the scaling functions $f_{gg}^{(k,0)}(\eta)$ with
$k=0,1$ in the $\overline {\rm MS}$ scheme for the exact and threshold
expressions (from \cite{betal}). We also show $f_{gg}^{(\pi^2),\,\rm res}$
which is defined in analogy to (2.11).
We see that the Born and first-order threshold
approximations are good only very close to threshold. The resummed result
enhances the Born cross section and tends to a positive constant at
threshold. The approximation is not good in the region $0.1<\eta<1$ so we
turn now to a discussion of the important terms in this region.

In \cite{LSN} an approximation was given for the
NLO soft-plus-virtual (S+V) contributions
and the analogy with the Drell-Yan process was exploited
to resum them to all orders of perturbation theory. The
S+V approximation is adequate in the kinematical region of
interest $0.1<\eta<1$ for the $q \bar q$ channel,
but not as good for the $gg$ channel in the $\overline {\rm MS}$ scheme.
Therefore we  reexamined the approximate formulae given in \cite{MSSN}
for the initial state gluon bremsstrahlung (ISGB) mechanism
to see if there are subleading terms that will improve the
S+V approximation. Let us see the structure of these terms.
We are discussing partonic reactions of the type
$i(k_1)+j(k_2) \rightarrow Q(p_1) + \bar Q(p_2)+g(k_3)$,
and we introduce the kinematic variables $t_1=(k_2-p_2)^2-m^2$,
$u_1=(k_1-p_2)^2-m^2$, and $s_4 = s+t_1+u_1$. The variable $s_4$ depends
on the four-momentum of the extra partons  emitted in the reaction.
The first-order S+V result for the
$q \bar {q}$ channel in the $\overline{\rm MS}$ scheme is
\begin{eqnarray}
s^2\frac{d^2\sigma^{(1)}_{q \bar q}(s,t_1,u_1)}{dt_1 du_1}&=&
\sigma_{q \bar q}^B(s,t_1,u_1)\frac{2C_F}{\pi} \alpha_s(\mu^2)
\nonumber \\ &&
\times
\left\{\left[\frac{1}{s_4}\left(2\ln\frac{s_4}{m^2}+\ln\frac{m^2}{\mu^2}
\right)\theta(s_4-\Delta) \right.\right.\nonumber \\ &&
\left.+\left(\ln^2\frac{\Delta}{m^2}
+\ln\frac{\Delta}{m^2}
\ln\frac{m^2}{\mu^2}\right) \delta(s_4)\right]
  \nonumber \\ &&
+\left.\left[-\frac{C_A}{2C_F}\frac{1}{s_4}\theta(s_4-\Delta)
-\frac{C_A}{2C_F}\ln\frac{\Delta}{m^2}
\delta(s_4)\right]\right\} \, \, \, \,
\end{eqnarray}
where
\begin{equation}
\sigma^B_{q\bar q}(s,t_1,u_1) = \pi \alpha_s^2(\mu^2) K_{q\bar{q}}
NC_F \Big[ \frac{t_1^2 + u_1^2}{s^2} + \frac{2m^2}{s}\Big]\,.
\end{equation}
Here $\Delta$ is a small parameter used to allow us to distinguish between
the soft $(s_4<\Delta)$ and the hard $(s_4>\Delta)$ regions in phase space.
The terms in the first pair of square brackets in (2.22) are the leading
S+V terms given in \cite{LSN} and those in the second pair of square
brackets are subleading terms that we want to examine.
In fig. 4 we plot the scaling functions $f_{q \bar q}^{(1,0)}$
for the exact result, the leading S+V result, and the S+V result with
both leading and subleading terms. The leading S+V result is
a reasonable approximation to the exact result in our region of interest
$0.1<\eta<1$.
The addition of Coulomb terms
worsens the leading S+V result. We also see that when we include the
subleading terms our approximation does not improve much in the region of
interest. However, when we add both the first order Coulomb term and
subleading terms to the leading
S+V result we get a very good agreement with the exact result.
Nevertheless, this is still not a major improvement over the
simple leading S+V result.
In the DIS scheme the analogous results are
\begin{eqnarray}
s^2\frac{d^2\sigma^{(1)}_{q \bar q}}{dt_1 du_1}(s,t_1,u_1)&=&
\sigma_{q \bar q}^B(s,t_1,u_1)
\frac{2 C_F}{\pi}\alpha_s(\mu^2)
 \nonumber \\ &&
\times
\left\{\left[\frac{1}{s_4}\left(\ln\frac{s_4}{m^2}+\ln\frac{m^2}{\mu^2}
\right)\theta(s_4-\Delta) \right.\right. \nonumber \\ &&
\left.+\left(\frac{1}{2}\ln^2\frac{\Delta}{m^2}+\ln\frac{\Delta}{m^2}
\ln\frac{m^2}{\mu^2}\right) \delta(s_4)\right]
 \nonumber \\ &&
+\left[\left(\frac{3}{4}+\ln2-\frac{C_A}{2C_F}\right)\frac{1}{s_4}
\theta(s_4-\Delta)\right.\nonumber \\ &&
\left.\left.+\left(\frac{3}{4}+\ln2-\frac{C_A}{2C_F}\right)
\ln\frac{\Delta}{m^2}\delta(s_4)\right]\right\}.
\end{eqnarray}
In fig. 5 we plot the corresponding scaling functions. Here
the addition of subleading terms worsens the leading S+V
approximation. The addition of Coulomb terms and large constants
enhances the first-order approximate results considerably. We also
show, for comparison, the results of the addition of the Coulomb
terms and the $\pi^2/3$
constant term only to the approximate results. These last curves
are the best fits to the exact result in the region $0.1<\eta<1$.

The resummation of the leading S+V terms has been given in \cite{LSN}.
The result is
\begin{eqnarray}
s^2\frac{d^2\sigma^{\rm res}_{q \bar q}
(s,t_1,u_1)}{dt_1 du_1}&=&\sigma_{q \bar q}^B(s,t_1,u_1)
\left[\frac{df(s_4/m^2,m^2/\mu^2)}{ds_4}\theta(s_4-\Delta)\right.
\nonumber \\ && \quad \quad \quad \quad \quad \quad
+\left.f(\frac{\Delta}{m^2},\frac{m^2}{\mu^2})\delta(s_4)\right],
\end{eqnarray}
where
\begin{equation}
f\left(\frac{s_4}{m^2},\frac{m^2}{\mu^2}\right)=
\exp\left[A\frac{C_F}{\pi}\bar\alpha_s\left(\frac{s_4}{m^2},m^2\right)
\ln^2\frac{s_4}{m^2}\right]
\frac{[s_4/m^2]^{\eta}}{\Gamma(1+\eta)}\exp(-\eta \gamma_E).
\end{equation}
Expressions for $A$, $\bar\alpha_s$, $\eta$, and $\gamma_E$ are
given in \cite{LSN}.
As the NNLO cross section is not known exactly
we do not how to resum the subleading terms. A reasonable guess
would be (2.25) with the function $f$ given now by
\begin{equation}
f\left(\frac{s_4}{m^2},\frac{m^2}{\mu^2}\right)=
f_{\rm Leading}\left(\frac{s_4}{m^2},\frac{m^2}{\mu^2}\right)
\exp\left[-\frac{C_A}{\pi}\alpha_s(\mu^2)\ln\frac{s_4}{m^2}\right]
\end{equation}
in the $\overline {\rm MS}$ scheme, and
\begin{equation}
f\left(\frac{s_4}{m^2},\frac{m^2}{\mu^2}\right)=
f_{\rm Leading}\left(\frac{s_4}{m^2},\frac{m^2}{\mu^2}\right)
\exp\left[\frac{C_F}{2\pi}\left(3\!+\!4\ln2-\frac{2C_A}{C_F}\right)
\alpha_s(\mu^2)
\ln\frac{s_4}{m^2}\right]
\end{equation}
in the DIS scheme, where now we call $f_{\rm Leading}$ the expression in
(2.26).

Now let us see the analogous results for the $gg$ channel in the
$\overline{\rm MS}$ scheme. We have
\begin{eqnarray}
s^2\frac{d^2\sigma^{(1)}_{gg}(s,t_1,u_1)}{dt_1 du_1}
&=&\sigma_{gg}^B(s,t_1,u_1)
\frac{2 C_A}{\pi}\alpha_s(\mu^2)
\nonumber \\ &&
\times\left\{\left[\frac{1}{s_4}
\left(2\ln\frac{s_4}{m^2}+\ln\frac{m^2}{\mu^2}\right)\theta(s_4-\Delta)
\right.\right.
\nonumber \\ &&
\left.\quad\quad+\delta(s_4)\left(\ln^2\frac{\Delta}{m^2}+\ln\frac{\Delta}{m^2}
\ln\frac{m^2}{\mu^2}\right)\right]
\nonumber \\ &&
+\left.\left[\frac{3C_A-8C_F}{-2C_A+8C_F}\left(\frac{1}{s_4}\theta(s_4-\Delta)
+\ln\frac{\Delta}{m^2}\delta(s_4)\right)\right]\right\}, \nonumber\\ &&
\end{eqnarray}
where
\begin{eqnarray}
\sigma^B_{gg}(s,t_1,u_1)& =&  2\pi \alpha_s^2(\mu^2) K_{gg}
NC_F \Big[C_F - C_A \frac{t_1u_1}{s^2}\Big] \nonumber \\ &&
\times\Big[ \frac{t_1}{u_1} + \frac{u_1}{t_1} + \frac{4m^2s}{t_1u_1}
\Big(1 - \frac{m^2s}{t_1u_1}\Big) \Big] \,.
\end{eqnarray}
Again, the terms in the first pair of square brackets in (2.29) are the
leading S+V terms and those in the second pair of square brackets
are subleading terms.
In fig. 6 we plot the scaling functions $f_{gg}^{(1,0)}$
for the exact result, the leading S+V result, and the S+V result with
both leading and subleading terms. We note that the leading
S+V approximate result is significantly smaller than the exact result
and that the addition of subleading terms improves the approximation
considerably. This is important since, as we will see in the next section,
the $gg$ channel is dominant for the production of $b$-quarks at HERA-B.
Also the addition of Coulomb terms further improves the approximation.
The resummation of the leading S+V terms for the $gg$ channel has
also been given in \cite{LSN}. The result is
\begin{eqnarray}
s^2\frac{d^2\sigma^{\rm res}_{gg}(s,t_1,u_1)}
{dt_1 du_1}&=&\sigma_{gg}^B(s,t_1,u_1)
\left[\frac{df(s_4/m^2,m^2/\mu^2)}{ds_4}\theta(s_4-\Delta)\right.
\nonumber \\ && \quad \quad \quad \quad \quad \quad
+\left.f\left(\frac{\Delta}{m^2},\frac{m^2}{\mu^2}\right)\delta(s_4)\right],
\end{eqnarray}
where
\begin{equation}
f\left(\frac{s_4}{m^2},\frac{m^2}{\mu^2}\right)=
\exp\left[2\frac{C_A}{\pi}\bar\alpha_s\left(\frac{s_4}{m^2},m^2\right)
\ln^2\frac{s_4}{m^2}\right]
\frac{[s_4/m^2]^{\eta}}{\Gamma(1+\eta)}\exp(-\eta \gamma_E).
\end{equation}
Again, as the NNLO cross section is not known exactly
we do not know how to resum the subleading terms. A reasonable guess
would be (2.31) with the function $f$ given now by
\begin{equation}
f\left(\frac{s_4}{m^2},\frac{m^2}{\mu^2}\right)=
f_{\rm Leading}\left(\frac{s_4}{m^2},\frac{m^2}{\mu^2}\right)
\exp\left[\frac{2C_A}{\pi}\frac{3C_A-8C_F}{-2C_A+8C_F}
\alpha_s(\mu^2)\ln\frac{s_4}{m^2}\right],
\end{equation}
where now we call $f_{\rm Leading}$ the expression in (2.32).

%------------------This is Section 3---------------------------------
\mysection{Results for bottom quark production at \newline
fixed-target $pp$ experiments and HERA-B}
%------------------------------------------------------------------
In this section we discuss $b$-quark production at HERA-B and also
at fixed-target $pp$ experiments in general,
and we examine the effects of the various
resummation procedures that were discussed in the previous section.
Following the notation in \cite{LSN} the total hadron-hadron
cross section in order $\alpha_s^{k}$ is
%--(4.1)
\begin{equation}
\sigma^{(k)}_H(S,m^2) = \sum_{ij}\int_{4m^2/S}^1
\,d\tau \,\Phi_{ij}(\tau,\mu^2)\, \sigma_{ij}^{(k)}(\tau S,m^2,\mu^2)\,,
\end{equation}
where $S$ is the square of the hadron-hadron c.m. energy and
$i,j$ run over $q,\bar q$ and $g$.
The parton flux $\Phi_{ij}(\tau,\mu^2)$ is defined via
%--(4.2)
\begin{equation}
\Phi_{ij}(\tau,\mu^2) = \int_{\tau}^1\, \frac{dx}{x}
H_{ij}(x,\frac{\tau}{x},\mu^2) \,,
\end{equation}
and $H_{ij}$ is a product of the scale-dependent parton distribution
functions $f^h_i(x,\mu^2)$, where $h$ stands for the hadron which is
the source of the parton $i$
%--(4.3)
\begin{equation}
H_{ij}(x_1, x_2, \mu^2) = f_i^{h_1}(x_1, \mu^2) f_j^{h_2}(x_2,\mu^2)\,.
\end{equation}
The mass factorization scale $\mu$ is chosen to be identical with
the renormalization scale in the running coupling constant.

In the case of the all-order
resummed expression the lower boundary in (3.1)
has to be modified according to the condition
$s_0 < s - 2ms^{1/2}$, where $s_0$ is defined below (see \cite{LSN}).
Resumming the soft gluon contributions to all orders we obtain
%--(4.4)
\begin{equation}
\sigma^{\rm res }_H(S,m^2) = \sum_{ij}\int_{\tau_0}^1
\,d\tau \,\Phi_{ij}(\tau,\mu^2)\, \sigma_{ij}(\tau S,m^2,\mu^2)\,,
\end{equation}
where $\sigma_{ij}$ is given in (3.24) of \cite{LSN} and
%--(4.6)
\begin{equation}
\tau_0 = \frac{[m+(m^2+s_0)^{1/2}]^2}{S}\,,
\end{equation}
with $s_0=m^2(\mu_0^2/\mu^2)^{3/2}$ ($\overline{\rm MS}$ scheme) or
$s_0=m^2(\mu_0^2/\mu^2)$  (DIS scheme). Here $\mu_0$ is the non-perturbative
parameter used in \cite{LSN}. It is used to cut off the resummation
since the resummed corrections diverge for small $\mu_0$.

We now specialize to bottom quark production at HERA-B
where $\sqrt{S}=39.2$ GeV.
In the presentation of our results for the exact, approximate,
and resummed hadronic cross sections
we use the MRSD$\_ ' \:$ parametrization for the parton distributions
\cite{mrs}.
Note that the hadronic results only involve partonic distribution
functions at moderate and large $x$, where there is little difference
between the various sets of parton densities.
We have used the MRSD$\_ '\:$ set 34 as given in PDFLIB \cite{PDFLIB} in the
DIS scheme with the number of active light flavors $n_f=4$ and the QCD
scale $\Lambda_5=0.1559$ GeV. We have used the two-loop corrected
running coupling constant as given by PDFLIB.

First, we discuss the NLO contributions to bottom quark production at HERA-B
using the results in [7-9].
Except when explicitly stated otherwise we will take the factorization
scale $\mu=m_b$ where $m_b$ is the $b$-quark mass.
Also, throughout the rest of this paper, we will use $m$ and $m_b$
interchangeably.
In fig. 7 we show the relative contributions of the $q \bar q$ channel in
the DIS scheme and the $gg$ channel in the $\overline{\rm MS}$ scheme
as a function
of the bottom quark mass. We see  that the $gg$ contribution is the dominant
one, lying between 70\% and 80\% of the total NLO cross section
for the range of bottom mass values given.
The $q \bar q$ contribution is smaller and makes up most of the remaining
cross section.
The relative contributions of the $g q$ and the $g \bar q$ channels in the
DIS scheme are negative and very small and they are also shown in the plot.
The situation here is the reverse of what is known about top quark production
at the Fermilab Tevatron where $q \bar q$ is the dominant channel
with $gg$ making up
the remainder of the cross section, and $g q$ and  $g \bar q$ making an even
smaller relative contribution than is the case for bottom quark production
at HERA-B. The reason for this difference between top quark and bottom quark
production is that the Tevatron is a $p \bar p$ collider while HERA-B is a
fixed-target $pp$ experiment. Thus, the parton densities involved
are different and since
sea quark densities are much smaller than valence quark densities, the
$q \bar q$ contribution to the hadronic cross section diminishes for a
fixed-target $pp$ experiment
relative to a $p \bar p$ collider for the same partonic cross section.

In fig. 8 we show the $K$ factors for the $q \bar q$ and $gg$ channels and for
their sum as a function of bottom quark mass.
The $K$ factor is defined by $K=(\sigma^{(0)}
+\sigma^{(1)}\mid _{\rm exact})/\sigma^{(0)}$, where $\sigma^{(0)}$ is the Born
term and $\sigma^{(1)}\mid _{\rm exact}$ is the exact first order correction.
We notice that the $K$ factor is quite large for the $gg$ channel,
which means that higher order effects are more important for this channel
than for $q \bar q$. Since $gg$ is the more important channel numerically,
the $K$ factor for the sum of the two channels is also quite large.
We also show the $K$ factor for the total which is slightly lower since
we are also taking into account the
negative contributions of the $qg$ and $\bar q g$ channels.

These large corrections come predominantly from the threshold region for
bottom quark production where it has been shown that initial state gluon
bremsstrahlung (ISGB) is responsible for the large corrections at
NLO \cite{MSSN}.
This can easily be seen in fig. 9 where the Born term and the
$O(\alpha_s^3)$ cross section are plotted as a function of
$\eta_{\rm cut}$ for the
$q \bar q$ and $gg$ channels, where
$\eta=(s -4m^2)/4m^2$ is the variable into which we have incorporated
the cut in our programs for the cross sections.
As we increase $\eta_{\rm cut}$ the
cross sections increase. The cross sections rise sharply for values of
$\eta_{\rm cut}$ between 0.1 and 1 and they reach a plateau at higher values
of $\eta_{\rm cut}$ indicating that the threshold region is very important
and that the region where $s>> 4m^2$ only makes a small contribution
to the cross sections.  This is the reason why we stressed in section 2
that our region of interest for comparison of the various approximations
at the partonic level was $0.1 <\eta <1$. In fig. 10 we plot as
a function of $\eta_{\rm cut}$ the Born term and the NLO
 cross section for the sum of the $q \bar q$ and $gg$ channels and also
the NLO cross section for the sum of all channels, including
the small negative contribution of the $qg$ and $\bar q g$ channels.
We thus see that the $qg$ and $\bar q g$
channels contribute a small negative contribution to the total exact NLO
cross section. Note that in the last two figures as well as throughout the
rest of this paper we are assuming that the bottom quark mass is
$m_b=4.75$ GeV$/c^2$.

Next, we discuss the scale dependence of our NLO results. In figs. 11 and 12
we show the Born term, the exact first-order correction, and the total
$O(\alpha_s^3)$ cross section as a function of the
factorization scale for the $q \bar q$ and $gg$ channels.
We see that as the scale decreases, the Born cross section increases without
bound but the exact first order correction decreases faster so that
the NLO cross section peaks at a scale close to half the mass of the
bottom quark and then decreases for smaller values of the scale. For the
$q \bar q$ channel the NLO cross section is relatively flat. The situation
is much worse for the $gg$ channel, however, since the peak is very sharp and
the scale dependence is much greater. Since the $gg$ channel dominates, this
large scale dependence is also reflected in the total cross section.
Thus the variation in
the NLO cross section for scales between $m/2$ and $2m$ is large. For
comparison we note that the scale dependence for top quark
production at the Fermilab Tevatron for $m_{\rm top}=175$ GeV $/c^2$
is much smaller.

In fig. 13 we plot the Born contribution for $\mu=m$ and the NLO
cross section for $\mu=m/2$, $m$, and $2m$,
as a function of the beam momentum for $b$-quark production at fixed-target
$pp$ experiments.
The big width of the
band reflects the large scale dependence that we discussed above.
We see that the NLO  cross section is almost twice as big
as the Born term for the whole range of beam momenta that we
are showing, and in particular for 820 GeV$/c$ which is the value
of the beam momentum at HERA-B.
The total NLO cross section for $b$-quark production at HERA-B is 28.8 nb for
$\mu=m/2$; 9.6 nb for $\mu=m$; and 4.2 nb for $\mu=2m$.
We also give the NLO results for the individual channels in fig. 14
for $\mu=m$.

In figs. 15 and 16 we examine the $\mu_0$ dependence of the
resummed cross section for $b$-quark production at HERA-B.
We also show, for comparison, the $\mu_0$ dependence of
$\sigma^{(0)}+\sigma^{(1)}\mid _{\rm app}+\sigma^{(2)}\mid _{\rm app}$
where we have imposed the
same cut on the phase space of $s_4$ ($s_4>s_0$)
as for the resummed cross section. Here $\sigma^{(1)}\mid _{\rm app}$
and $\sigma^{(2)}\mid _{\rm app}$ denote the approximate first and second
order corrections, respectively, where only soft gluon contributions
are taken into account.
The effect of the resummation shows in the difference between the two curves.
At small $\mu_0$, $\sigma^{\rm res}$ diverges, signalling the
presence of the infrared renormalon.
There is a region where the higher-order terms are numerically
important, for instance $0.5<\mu_0<1$ in fig. 15.
At high values of $\mu_0$ the two lines are practically the same.
For the $q \bar q$ channel in the DIS scheme the resummation is successful
in the sense that there is a relatively large region of $\mu_0$ where
resummation is well behaved before we encounter the divergence. For the
$gg$ channel, however, the situation is not as good.
{}From these curves we choose what we think are reasonable values for $\mu_0$.
We choose $\mu_0=0.6$ GeV for the $q \bar q$ channel
and $\mu_0=1.7$ GeV for the $gg$ channel. The value we chose for the $gg$
channel is such that the resummed cross section is a little bit higher than the
sum $\sigma^{(0)}+\sigma^{(1)}\mid _{\rm app}+\sigma^{(2)}\mid _{\rm app}$.

Using the values of $\mu_0$ that we chose from the previous graphs,
we proceed to plot the resummed cross section  for $b$-quark
production at fixed-target $pp$ experiments versus beam momentum.
We present the
results in fig. 17 for the $q \bar q$ and $gg$ channels.
We also show the results we have if in addition we resum subleading terms,
and also if we resum both subleading terms and Coulomb and constant terms.
For comparison the exact NLO results are shown as well. The
resummed cross sections were calculated with the cut $s_4>s_0$ while no
such cut was imposed on the NLO result.
Since we know the exact $O(\alpha_s^3)$ result, we can make an even
better estimate by calculating
the perturbation theory improved cross sections
defined by
\begin {equation}
\sigma_H^{\rm imp}=\sigma_H^{\rm res}
+\sigma_H^{(1)}\mid _{\rm exact}
-\sigma_H^{(1)}\mid _{\rm app}\,,
\end{equation}
to exploit the fact that $\sigma_H^{(1)}\mid _{\rm exact}$
is known and $\sigma_H^{(1)}\mid _{\rm app}$
is included in
$\sigma_H^{\rm res}$.
Then, in fig. 18 we plot the improved total cross section versus
beam momentum (where we have
also taken into account the small negative contributions of the $qg$ and
$\bar q g$ channels) and, for comparison, the total exact NLO cross section
for the three choices $\mu=m/2$, $m$, and $2m$.
We also show the improved total cross section including resummation of
subleading terms, and of both subleading terms and Coulomb and constant terms.
The last curve lies above the NLO result for $\mu=m/2$ for most values of the
beam momenta shown (including the value at HERA-B)
so that the effect of resummation exceeds the scale dependence of
the NLO cross section. The improved total cross section for
$b$-quark production at HERA-B is 19.4 nb, if we resum
leading terms only; 23.8 nb, if we resum leading and subleading terms;
and 31.4 nb, if we resum leading, subleading, Coulomb, and constant terms.

Finally, we present some results on the inclusive transverse momentum
($p_t$) and rapidity ($Y$) distributions of the bottom quark at HERA-B.
Heavy quark differential distributions are known in NLO \cite{bnmss,nde2}.
Some of the relevant
formulae for this part have been given already in \cite{NKJS},
where the $p_t$ and $Y$ distributions, including resummation,
were presented for top-quark
production at the Fermilab Tevatron.
The heavy-quark
inclusive differential distribution in $p_t^2$ is given by
\begin{equation}
\frac{d\sigma^{(k)}_H(S,m^2,p_t^2)}{dp_t^2} = \sum_{ij}\int_{4m_t^2/S}^1
\,d\tau \,\Phi_{ij}(\tau,\mu^2)\, \frac{d\sigma_{ij}^{(k)}
(\tau S,m^2,p_t^2,\mu^2)}{dp_t^2}\,,
\end{equation}
with $m_t^2 = m^2 + p_t^2$. In the case of the all-order
resummed expression the lower boundary in (3.7)
has to be modified according to the condition
$s_0 < s - 2m_ts^{1/2}$.
Resumming the soft gluon contributions to all orders we obtain
\begin{equation}
\frac{d\sigma^{\rm res}_H(S,m^2,p_t^2)}{dp_t^2} = \sum_{ij}\int_{\tau_0}^1
\,d\tau \, \Phi_{ij}(\tau,\mu^2) \frac{d\sigma_{ij}(\tau S,m^2,p_t^2,\mu^2)}
{dp_t^2}\,,
\end{equation}
with $d\sigma_{ij}/dp_t^2$ given in (3.6) of \cite{NKJS}
and
\begin{equation}
\tau_0 = \frac{[m_t+(m_t^2+s_0)^{1/2}]^2}{S}\,.
\end{equation}

The corresponding formula to (3.7) for the heavy quark inclusive
differential distribution in $Y$ is
\begin{equation}
\frac{d\sigma^{(k)}_H(S,m^2,y)}{dY} = \sum_{ij}\int_{4m^2\cosh^2 y/S}^1
\,d\tau \,\Phi_{ij}(\tau,\mu^2)\, \frac{d\sigma_{ij}^{(k)}
(\tau S,m^2,y,\mu^2)}{dy}\,.
\end{equation}
The all-order resummed differential distribution in $Y$ is given by
\begin{equation}
\frac{d\sigma^{\rm res}_H(S,m^2,Y)}{dY} = \sum_{ij}\int_{\tau_0}^1
\,d\tau \,\Phi_{ij}(\tau,\mu^2)\, \frac{d\sigma_{ij}(\tau S,m^2,y,\mu^2)}
{dy}\,,
\end{equation}
with $d\sigma_{ij}/dy$ given in (3.9) of \cite{NKJS}
and
\begin{equation}
\tau_0 = \frac{[m\cosh y +(m^2\cosh^2 y+s_0)^{1/2}]^2}{S}\,.
\end{equation}
The hadronic heavy quark rapidity $Y$ is related to the
partonic heavy quark rapidity $y$
by
\begin{equation}
Y=y+\frac{1}{2}\ln\frac{x_1}{x_2}\,.
\end{equation}

We begin with the $p_t$ distributions.
For these plots the mass factorization scale is not everywhere equal to $m$.
We chose $\mu=m$ in $s_0$, $f_k(s_4/m^2\,,m^2/\mu^2)$ and $\bar{\alpha}_s$,
but $\mu=m_t$ in the MRSD$\_ ' \:$ parton distribution
functions and the running coupling constant $\alpha_s(\mu)$.
In fig. 19, we give the results for the $q \bar q$ channel in the DIS scheme.
We plot the Born term $d\sigma_H^{(0)}/dp_t$, the first order exact result
$d\sigma_H^{(1)}/dp_t\mid _{\rm exact}$, the first order approximation
$d\sigma_H^{(1)}/dp_t\mid _{\rm app}$, the second
order approximation $d\sigma_H^{(2)}/dp_t\mid _{\rm app}$,
and the resummed result $d\sigma_H^{\rm res}/dp_t$ for $\mu_0=0.6$ GeV.
This is the same value for $\mu_0$ that was used above for the total cross
section. We also show resummed results with the inclusion of subleading terms,
and with both subleading terms and Coulomb and constant terms.
If we decrease $\mu_0$ the differential cross
sections will increase.
The resummed distributions
were calculated with the cut $s_4>s_0$ while no such cut was imposed
on the phase space for the individual terms in the perturbation series.
We continue with the results for the $gg$ channel in
the $\overline{\rm MS}$ scheme.
The corresponding plot is given in fig. 20.
We note that the corrections in this channel are large. In fact the exact
first-order correction is larger than the Born term and the approximate
second-order correction is larger than the approximate first-order correction.
In this case the value of $\mu_0$ has been chosen to be
$\mu_0=1.7$ GeV as above.
We define the improved $p_t$ distribution by
\begin {equation}
\frac{d\sigma_H^{\rm imp}}{dp_t}=\frac{d\sigma_H^{\rm res}}{dp_t}
+\frac{d\sigma_H^{(1)}}{dp_t}\mid _{\rm exact}
-\frac{d\sigma_H^{(1)}}{dp_t}\mid _{\rm app}\,.
\end{equation}
In fig. 21 we plot the improved $p_t$ distributions for the sum
of all channels,
where we have included the small negative contributions of the $qg$
and $\bar q g$ channels.
For comparison we also show the total exact NLO results for
$\mu=m_t/2$, $m_t$, and
$2m_t$. The improved $p_t$ distributions are uniformly above the exact
$O(\alpha_s^3)$ results.
We see that the effect of the resummation exceeds the uncertainty
due to scale dependence.

We finish with a discussion of the $Y$ distributions.
In this case we set the factorization mass scale equal to $m$ everywhere.
We begin with the $q \bar q$ channel. In fig. 22 we show
the Born term $d\sigma_H^{(0)}/dY$, the first order exact result
$d\sigma_H^{(1)}/dY\mid _{\rm exact}$, the first order approximation
$d\sigma_H^{(1)}/dY\mid _{\rm app}$, the second
order approximation $d\sigma_H^{(2)}/dY\mid _{\rm app}$,
and the resummed result $d\sigma_H^{\rm res}/dY$ for $\mu_0=0.6$ GeV.
We also show resummed results with the inclusion of subleading terms,
and with both subleading terms and Coulomb and constant terms.
Again, the resummed distributions were calculated with the cut $s_4>s_0$
while no such cut was imposed
on the phase space for the individual terms in the perturbation series.
We continue with the results for the $gg$ channel
in the $\overline{\rm MS}$ scheme.
The corresponding plot is given in fig. 23. Here, the value
of $\mu_0$ is $\mu_0=1.7$ GeV.
The corrections in this channel are large as was the case for the $p_t$
distributions.
We define the improved $Y$ disrtibution by
\begin {equation}
\frac{d\sigma_H^{\rm imp}}{dY}=\frac{d\sigma_H^{\rm res}}{dY}
+\frac{d\sigma_H^{(1)}}{dY}\mid _{\rm exact}
-\frac{d\sigma_H^{(1)}}{dY}\mid _{\rm app}\,.
\end{equation}
In fig. 24 we plot the improved $Y$ distributions for the sum
of all channels,
where we have included the small negative contributions of the $qg$
and $\bar q g$ channels.
For comparison we also show the total exact NLO results
for $\mu=m/2$, $m$, and
$2m$. The improved $Y$ distributions are uniformly above the $O(\alpha_s^3)$
results.
Again, we see that the effect of the resummation exceeds the uncertainty
due to scale dependence.

%------------------This is Section 4---------------------------------
\mysection{Conclusions}
%------------------------------------------------------------------
We have presented NLO and resummed results for the cross section
and differential
distributions for bottom quark production at HERA-B. Results for the
cross section as a function of beam momentum have also been given for
fixed-target $pp$ experiments in general.
It has been shown that the $gg$ channel is dominant and
that the threshold region gives the main contribution
to the NLO cross section. Approximations for the soft gluon contributions
in that region have been compared with the exact results.
The resummation of the leading S+V logarithms
produces an enhancement of the NLO results.
The leading S+V approximation is not very good in the $gg$ channel in the
$\overline {\rm MS}$ scheme in the kinematic region
that is important for bottom quark production at HERA-B. The addition
of subleading S+V terms and Coulomb terms improves the approximation
considerably. The resummation of these additional terms
further enhances the cross section. We must stress, however, that our
formula for the exponentiation of subleading terms is not based on
any rigorous analysis
and more work in this area will have to be done in the future.

{\bf ACKNOWLEDGEMENTS}

The work in this paper was supported in part under the
contract NSF 93-09888.

%----------------------------References-------------------------------------
%

\pagebreak
%------------------Figures---------------------------------
\centerline{\Large \bf List of Figures}
\vspace{3mm}

Fig. 1. The scaling functions $f_{q\bar q}^{(k,0)}$ in the $\overline {\rm MS}$
scheme. Plotted are $f_{q\bar q}^{(0,0)}$ (exact, upper solid line at large
$\eta$; threshold approximation, upper dotted line at large $\eta$),
$f_{q\bar q}^{(1,0)}$ (exact, lower solid line at large $\eta$;
threshold approximation, lower dotted line at large $\eta$), and
$f_{q\bar q}^{(\pi^2),\,\rm res}$ (dashed line).

Fig. 2. Same as fig. 1 but now for the DIS scheme.
Also shown is $f_{q\bar q}^{(\pi^2),\,\rm res}$ where the only
constant that we exponentiate is the $\pi^2/3$ term (dash-dotted line).

Fig. 3. Same as fig. 1 but now for $f_{gg}^{(k,0)}$ in the
$\overline {\rm MS}$ scheme.

Fig. 4. The scaling functions $f_{q\bar q}^{(1,0)}$ in the $\overline {\rm MS}$
scheme.
Plotted are the exact result (solid line), the leading S+V
approximation (dotted line),
the leading S+V approximation plus Coulomb terms (short-dashed line),
the S+V approximation with leading plus subleading terms (long dashed line),
and the S+V approximation with leading plus subleading terms
plus Coulomb terms (dash-dotted line).

Fig. 5. Same af fig. 4 but now for the DIS scheme. Also shown are
the leading S+V approximation plus Coulomb terms and the $\pi^2/3$
constant term only (lower short-dashed line),
and the S+V approximation with leading plus subleading terms
plus Coulomb terms and the $\pi^2/3$ constant term only
(lower dash-dotted line).

Fig. 6.  Same as fig. 4 but now for $f_{gg}^{(1,0)}$ in the
$\overline {\rm MS}$ scheme.

Fig. 7. Fractional contributions of the $gg$ ($\overline{\rm MS}$ scheme,
short-dashed line), $q \bar q$ (DIS scheme, long-dashed line),
$qg$ (DIS scheme, lower dotted line), and $\bar q g$ (DIS scheme,
upper dotted line) channels to the total $O(\alpha_s^3)$
$b$-quark production cross section at HERA-B as a function of $b$-quark mass.

Fig. 8. The K factors as a function of $b$-quark mass
for $b$-quark production at HERA-B for the
$gg$ channel($\overline{\rm MS}$ scheme,
short-dashed line), the $q \bar q$ channel (DIS scheme, long-dashed line),
the sum of the $gg$ and $q \bar q$ channels (dotted line), and the sum of all
channels (solid line).

Fig. 9. Cross sections for $b$-quark production at HERA-B
versus $\eta_{\rm cut}$ with $m_b=4.75$ GeV$/c^2$ for the $q\bar q$ channel
in the DIS scheme and the $gg$ channel in the $\overline{\rm MS}$ scheme.
Plotted are the Born term ($gg$, upper solid line at high $\eta_{\rm cut}$;
$q\bar q$, lower solid line at high $\eta_{\rm cut}$)
and the $O(\alpha_s^3)$ cross section ($gg$, upper dashed line;
$q\bar q$, lower dashed line).

Fig. 10. Cross sections for $b$-quark production at HERA-B
versus $\eta_{\rm cut}$ with $m_b=4.75$ GeV$/c^2$.
Plotted are the total Born term (solid line), the total $O(\alpha_s^3)$
cross section (dashed line), and the $O(\alpha_s^3)$ cross section for the
sum of the $q\bar q$ and $gg$ channels (dotted line).

Fig. 11. The scale dependence of the cross section for $b$-quark production
at HERA-B with $m_b=4.75$ GeV$/c^2$ for the $q \bar q$ channel
in the DIS scheme. Plotted are the Born term (solid line), the exact
first-order correction (dotted line),
and their sum (dashed line).

Fig. 12. Same as fig. 11 but now for the $gg$ channel
in the $\overline{\rm MS}$ scheme.

Fig. 13. The total Born (dotted line) and $O(\alpha_s^3)$
($\mu=m$ solid line, $\mu=m/2$ upper dashed line, and $\mu=2m$
lower dashed line)
$b$-quark production cross sections at fixed-target $pp$ experiments
versus beam momentum for $m_b=4.75$ GeV$/c^2$.

Fig. 14. Contributions of individual channels to the total
$O(\alpha_s^3)$ $b$-quark
production cross section at fixed-target $pp$ experiments
versus beam momentum for $m_b=4.75$ GeV$/c^2$.
Plotted are the contributions of the $gg$
($\overline {\rm MS}$ scheme, short-dashed
line) and $q \bar q$ (DIS scheme, long-dashed line) channels,
and the absolute value of
the contributions of the $qg$ (DIS scheme, upper dotted line) and $\bar q g$
(DIS scheme, lower dotted line) channels.

Fig. 15. The $\mu_0$ dependence of the resummed cross section for
$b$-quark production at HERA-B with $m_b=4.75$ GeV$/c^2$
for the $q \bar q$ channel in the DIS scheme.
Plotted are $\sigma_{q\bar q}^{\rm res}$ (solid line) and the sum
$\sigma^{(0)}+\sigma^{(1)}\mid _{\rm app}+\sigma^{(2)}\mid _{\rm app}$
(dotted line).

Fig. 16. Same as fig. 15 but now for the $gg$ channel in the
$\overline{\rm MS}$ scheme.

Fig. 17. Resummed and NLO cross sections versus beam momentum
for $b$-quark production at fixed-target $pp$ experiments for $m_b=4.75$
GeV$/c^2$. Plotted are the resummed cross sections for the
$q \bar q$ channel in the DIS scheme for $\mu_0=0.6$ GeV
(leading terms only, short-dashed line; with subleading terms,
lower short-dash-dotted line;
with both subleading terms and Coulomb and constant terms, upper
short-dash-dotted line)
and for the $gg$ channel in the $\overline{\rm MS}$ scheme for
$\mu_0=1.7$ GeV (leading terms only, long-dashed line;
with subleading terms, lower long-dash-dotted line;
with both subleading terms and Coulomb terms, upper
long-dash-dotted line); and the $O(\alpha_s^3)$
cross sections for the $gg$ channel in the $\overline{\rm MS}$ scheme
and the $q \bar q$ channel in the DIS scheme (upper and lower
solid lines, respectively).

Fig. 18.  Improved and NLO cross sections versus beam momentum
for $b$-quark production at fixed-target $pp$ experiments for $m_b=4.75$
GeV$/c^2$. Plotted are the total improved cross section
(leading terms only, short-dashed line;
with subleading terms, lower long-dashed line;
with both subleading terms and Coulomb and constant terms, upper
long-dashed line)
and the total $O(\alpha_s^3)$ result
($\mu=m$ solid line, $\mu=m/2$ upper dotted line, $\mu=2m$ lower
dotted line).

Fig. 19. The bottom quark $p_t$ distributions $d\sigma_H^{(k)}/dp_t$
at HERA-B for
the $q\bar{q}$ channel in the DIS scheme for
$m_b=4.75$ GeV$/c^2$.
Plotted are $d\sigma_H^{(0)}/dp_t$ (solid line),
$d\sigma_H^{(1)}/dp_t\mid _{\rm exact}$ (dotted line),
$d\sigma_H^{(1)}/dp_t\mid _{\rm app}$ (short-dashed line),
$d\sigma_H^{(2)}/dp_t\mid _{\rm app}$ (long-dashed line),
and $d\sigma_H^{\rm res}/dp_t$ for $\mu_0=0.6$ GeV
(leading terms only, short-dash-dotted line;
with subleading terms, lower long-dash-dotted line; with both subleading
terms and Coulomb and constant terms, upper long-dash-dotted line).

Fig. 20. Same as fig. 19 but now for the $gg$ channel in the
$\overline {\rm MS}$ scheme and with $\mu_0=1.7$ GeV.

Fig. 21. The bottom quark $p_t$ distributions  $d\sigma_H/dp_t$
at HERA-B for the sum
of all channels for  $m_b=4.75$ GeV$/c^2$.
Plotted are $d\sigma_H^{(0)}/dp_t+d\sigma_H^{(1)}/dp_t\mid _{\rm exact}$
($\mu=m_t$ solid line, $\mu=m_t/2$ upper dotted line, $\mu=2m_t$
lower dotted line)
and $d\sigma_H^{\rm imp}/dp_t$
(leading terms only, short-dashed line;
with subleading terms, lower long-dashed line; with both subleading
terms and Coulomb and constant terms, upper long-dashed line).

Fig. 22. The bottom quark $Y$ distributions $d\sigma_H^{(k)}/dY$
at HERA-B for
the $q\bar{q}$ channel in the DIS scheme for $m_b=4.75$ GeV$/c^2$.
Plotted are $d\sigma_H^{(0)}/dY$ (solid line),
$d\sigma_H^{(1)}/dY\mid _{\rm exact}$ (dotted line),
$d\sigma_H^{(1)}/dY\mid _{\rm app}$ (short-dashed line),
$d\sigma_H^{(2)}/dY\mid _{\rm app}$ (long-dashed line),
and $d\sigma_H^{\rm res}/dY$ for $\mu_0=0.6$ GeV
(leading terms only, short-dash-dotted line;
with subleading terms, lower long-dash-dotted line; with both subleading
terms and Coulomb and constant terms, upper long-dash-dotted line).

Fig. 23.  Same as fig. 22 but now for the $gg$ channel in the
$\overline {\rm MS}$ scheme and with $\mu_0=1.7$ GeV.

Fig. 24. The bottom quark $Y$ distributions  $d\sigma_H/dY$
at HERA-B for the sum of all channels for $m_b=4.75$ GeV$/c^2$.
Plotted are $d\sigma_H^{(0)}/dY+d\sigma_H^{(1)}/dY\mid _{\rm exact}$
($\mu=m$ solid line, $\mu=m/2$ upper dotted line, $\mu=2m$ lower dotted line)
and $d\sigma_H^{\rm imp}/dY$
(leading terms only, short-dashed line;
with subleading terms, lower long-dashed line; with both subleading
terms and Coulomb and constant terms, upper long-dashed line).

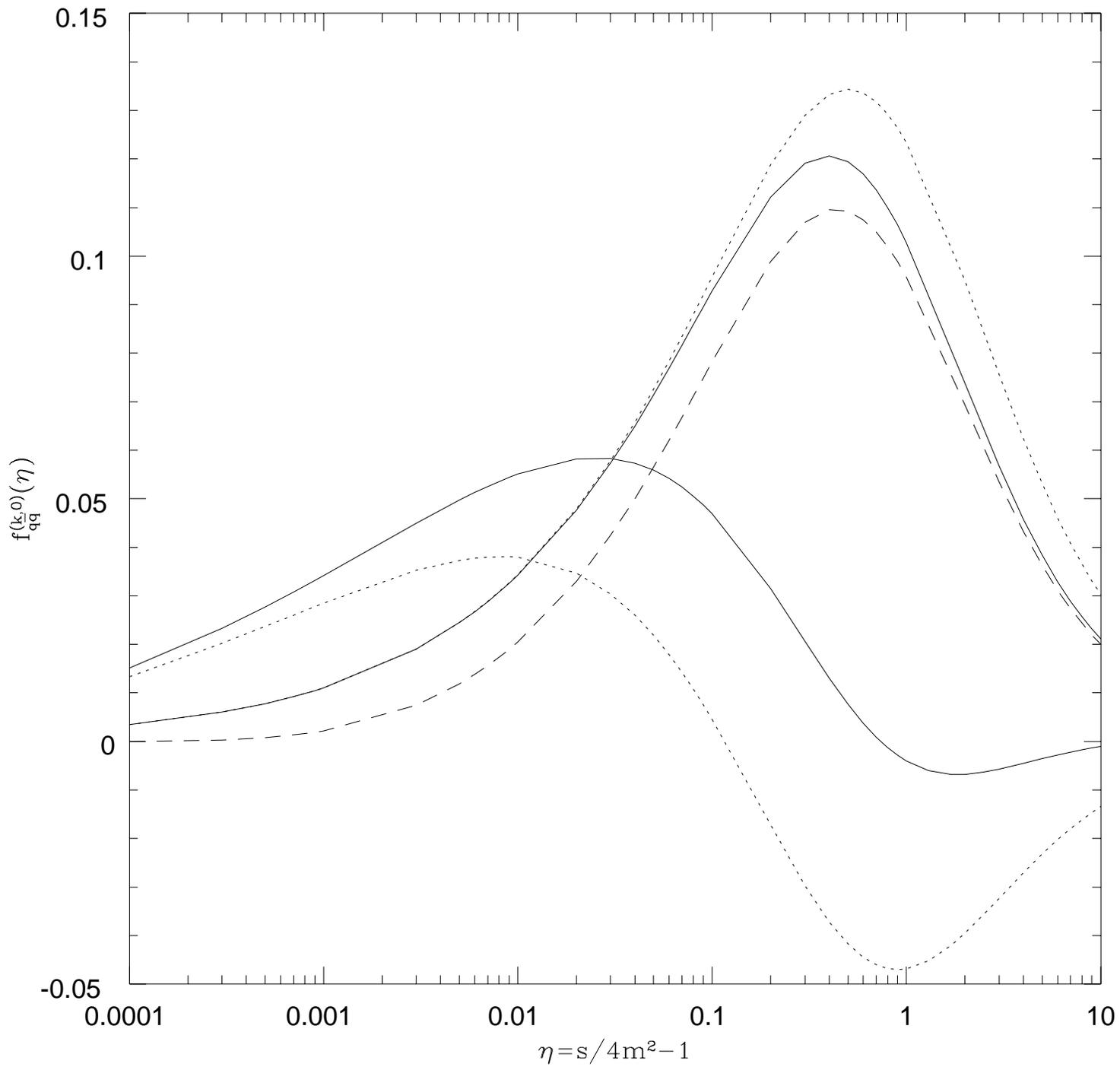
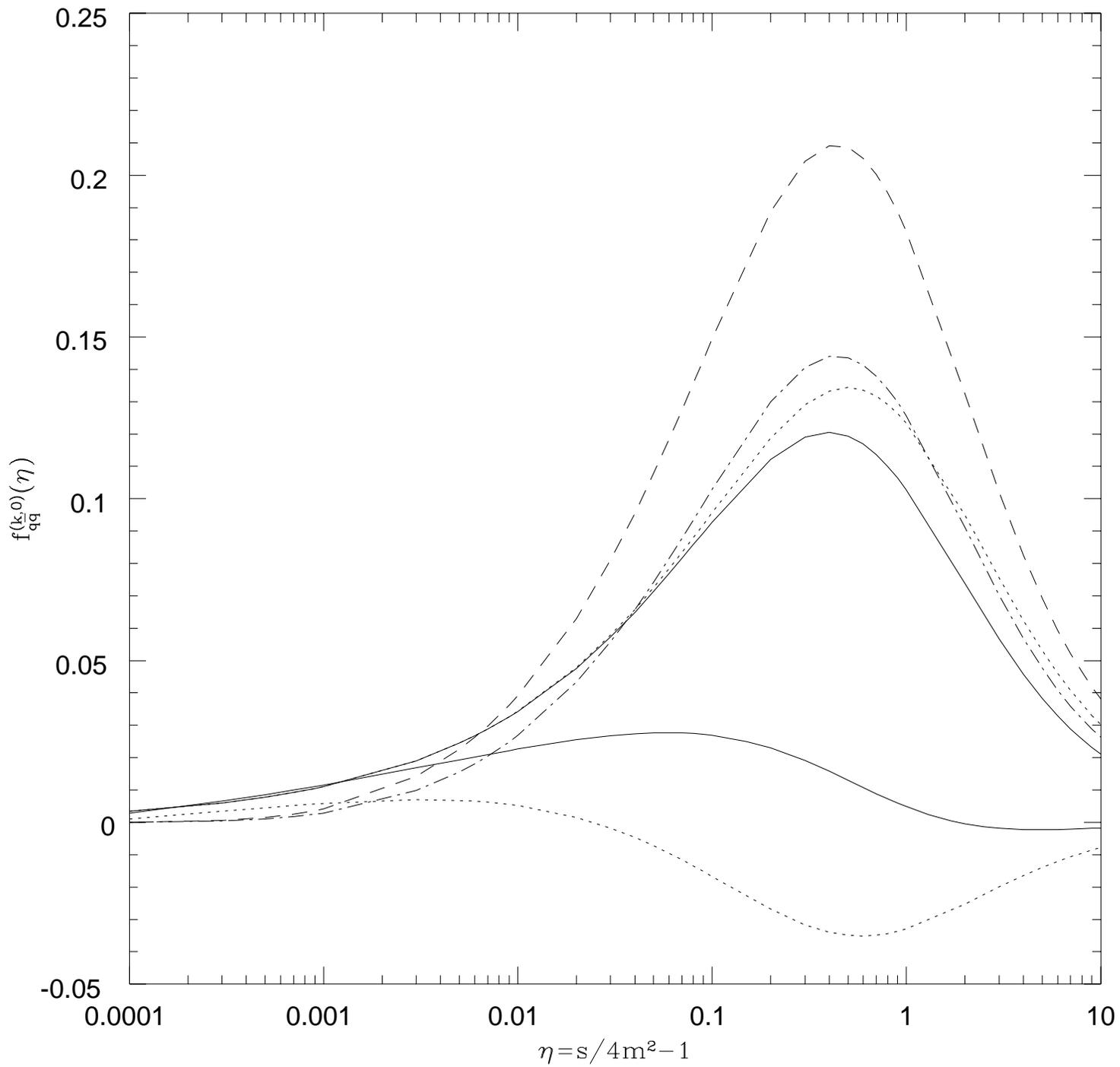
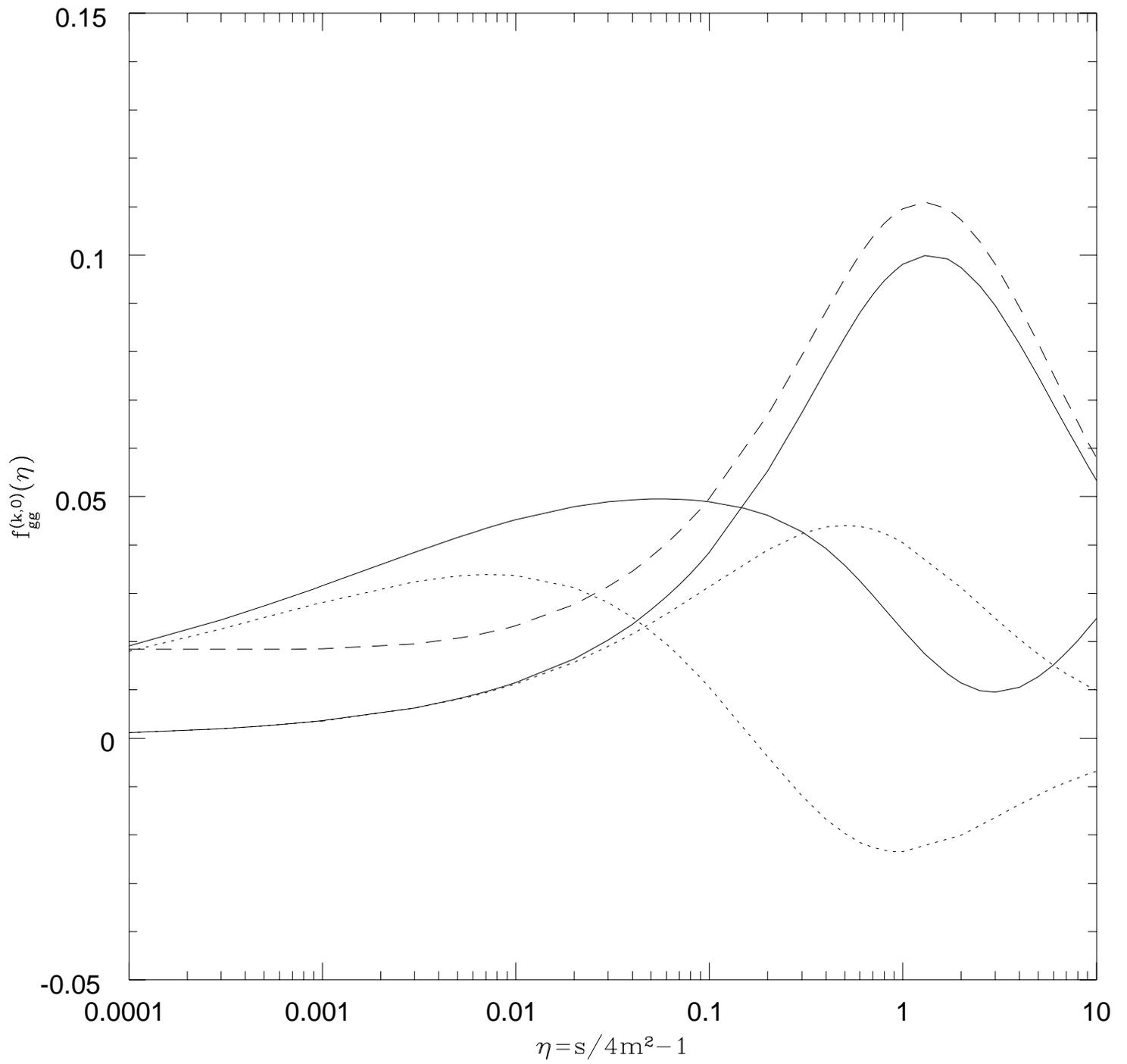
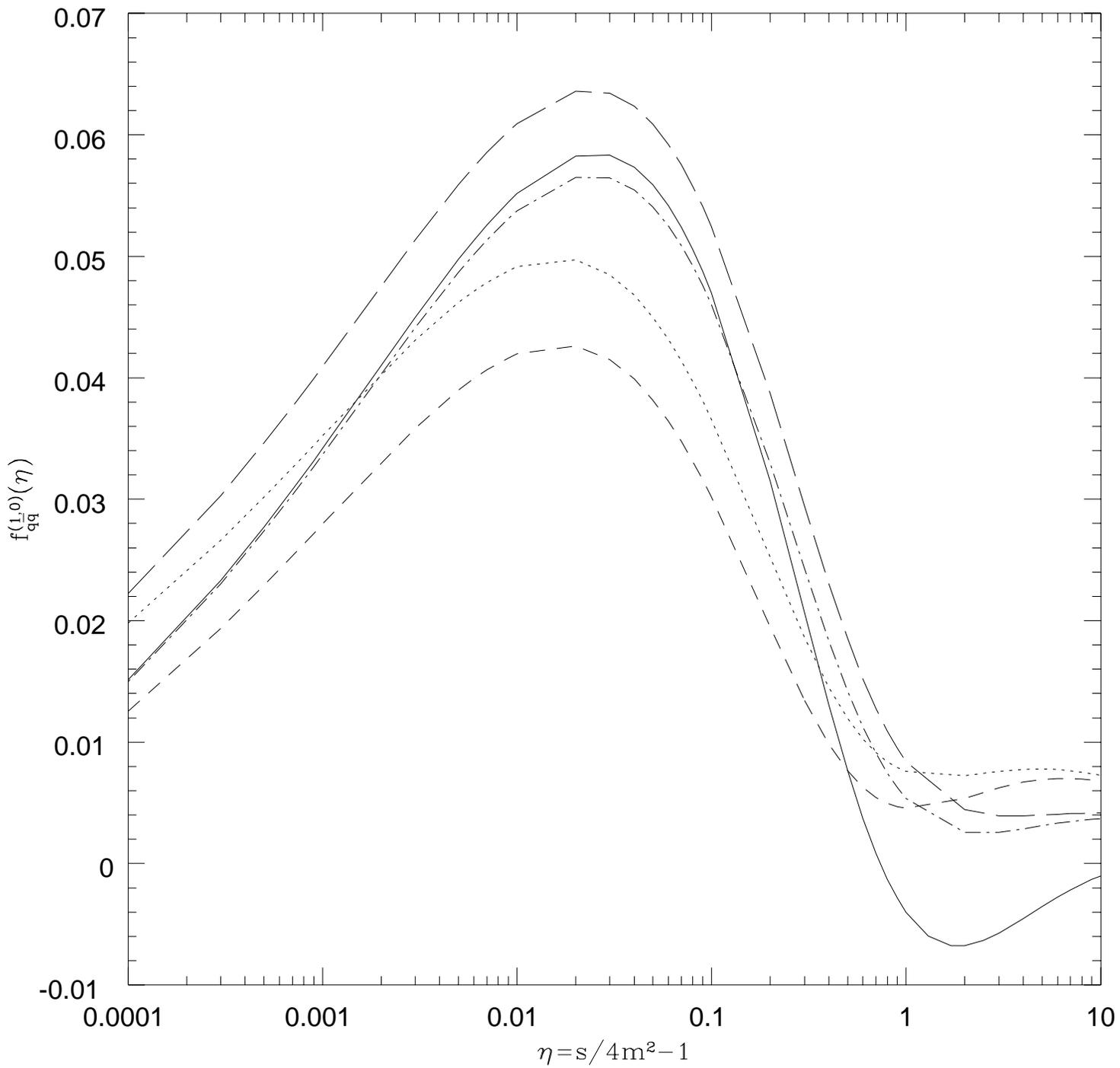
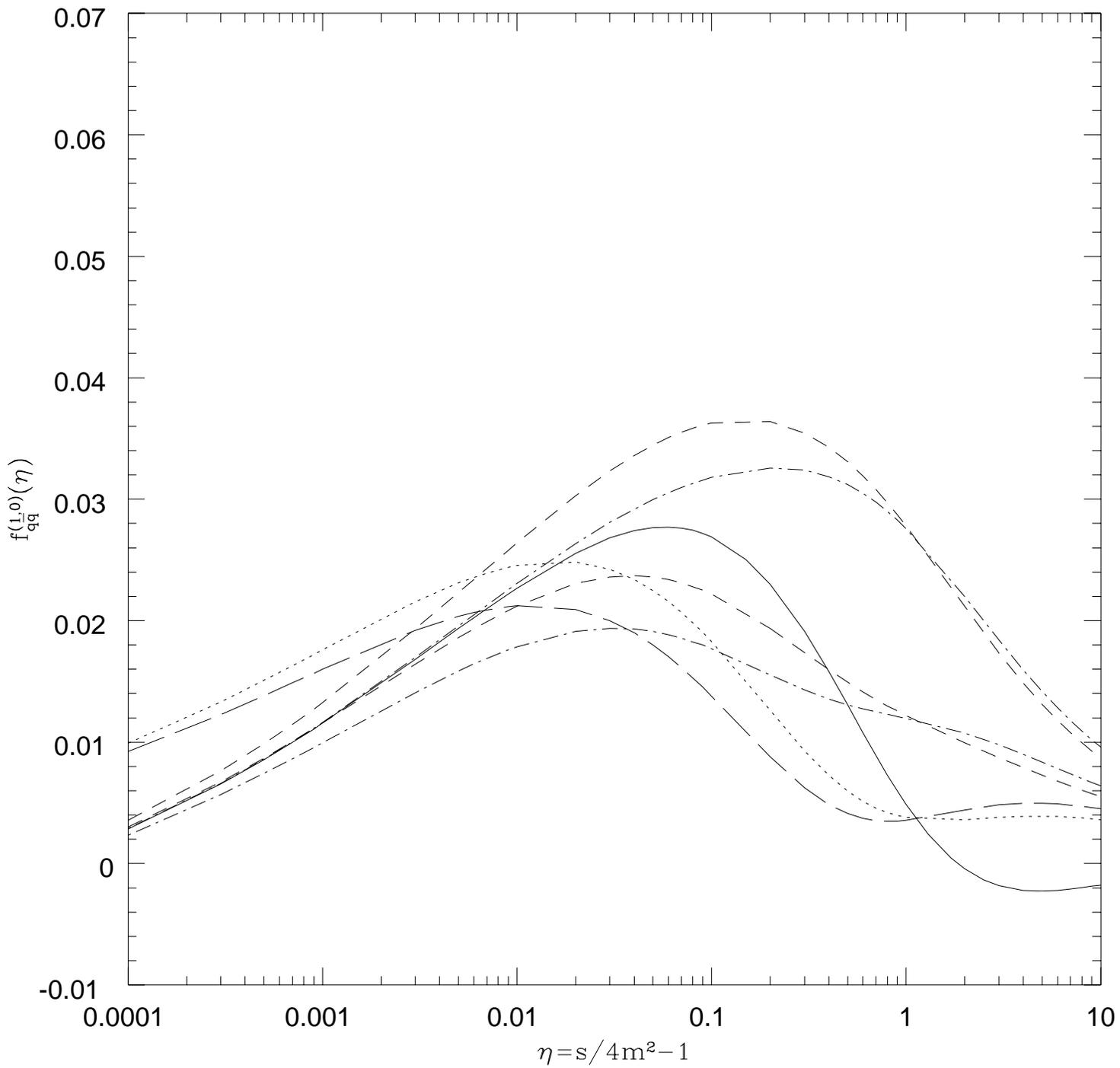
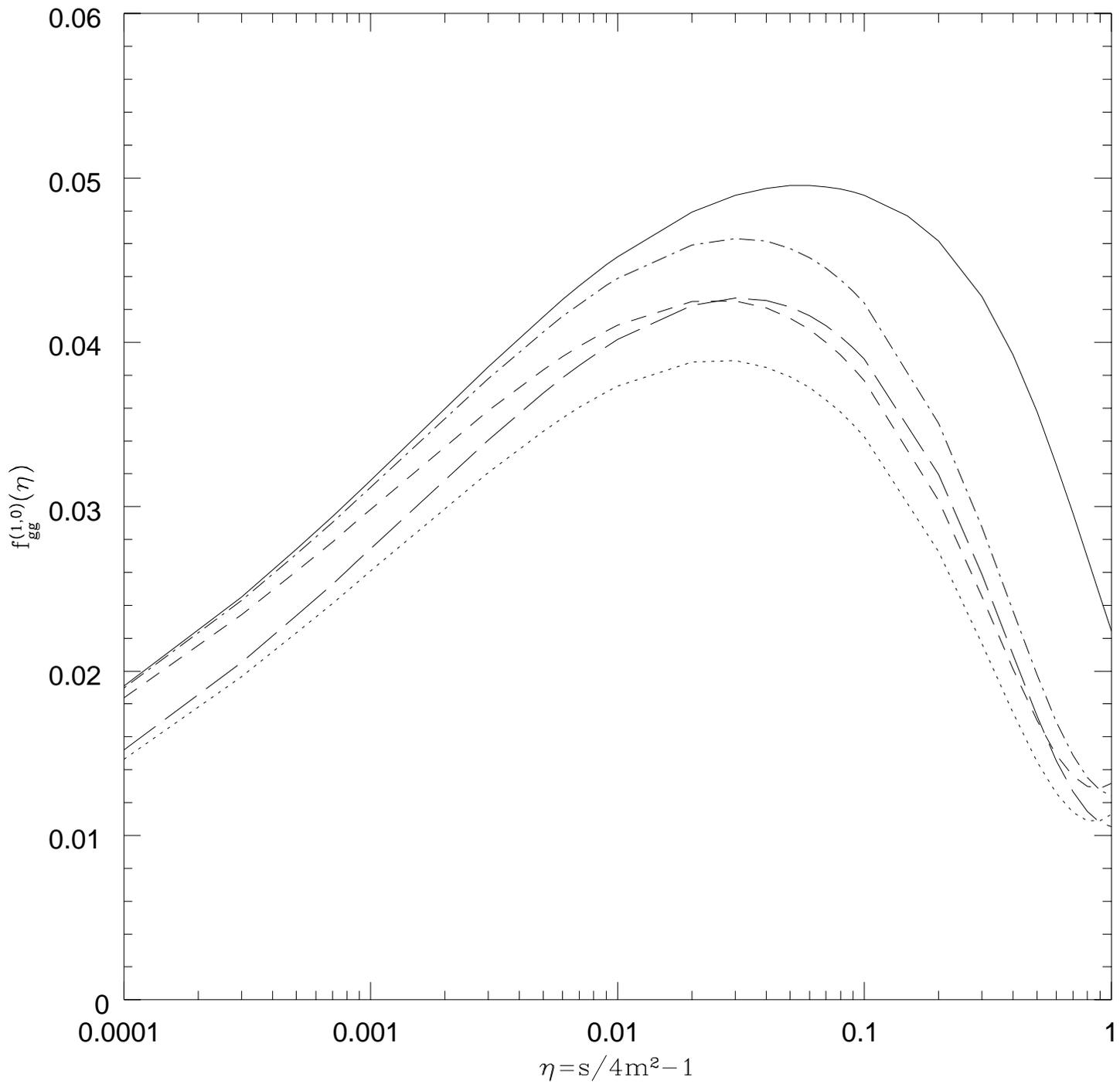
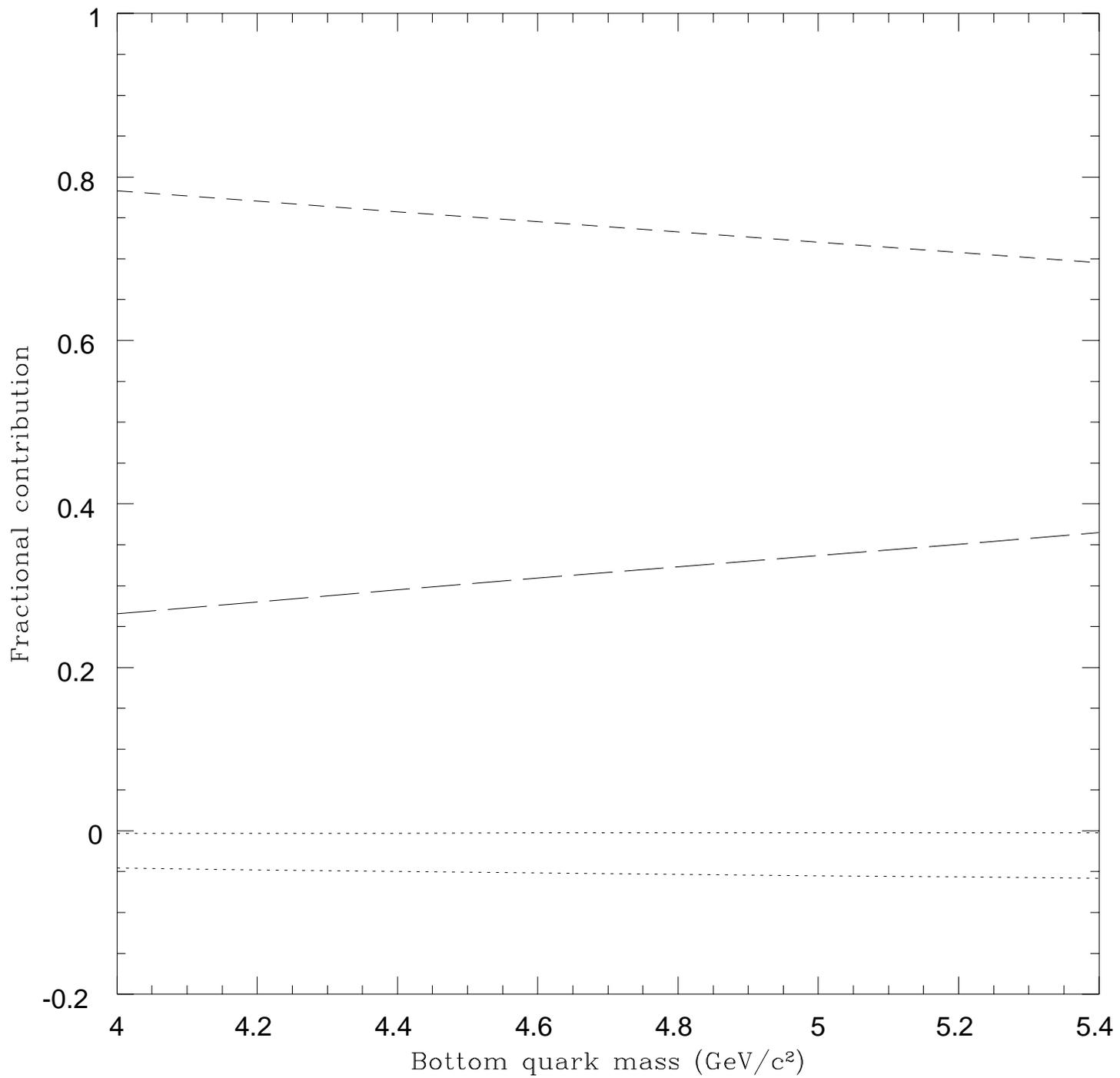
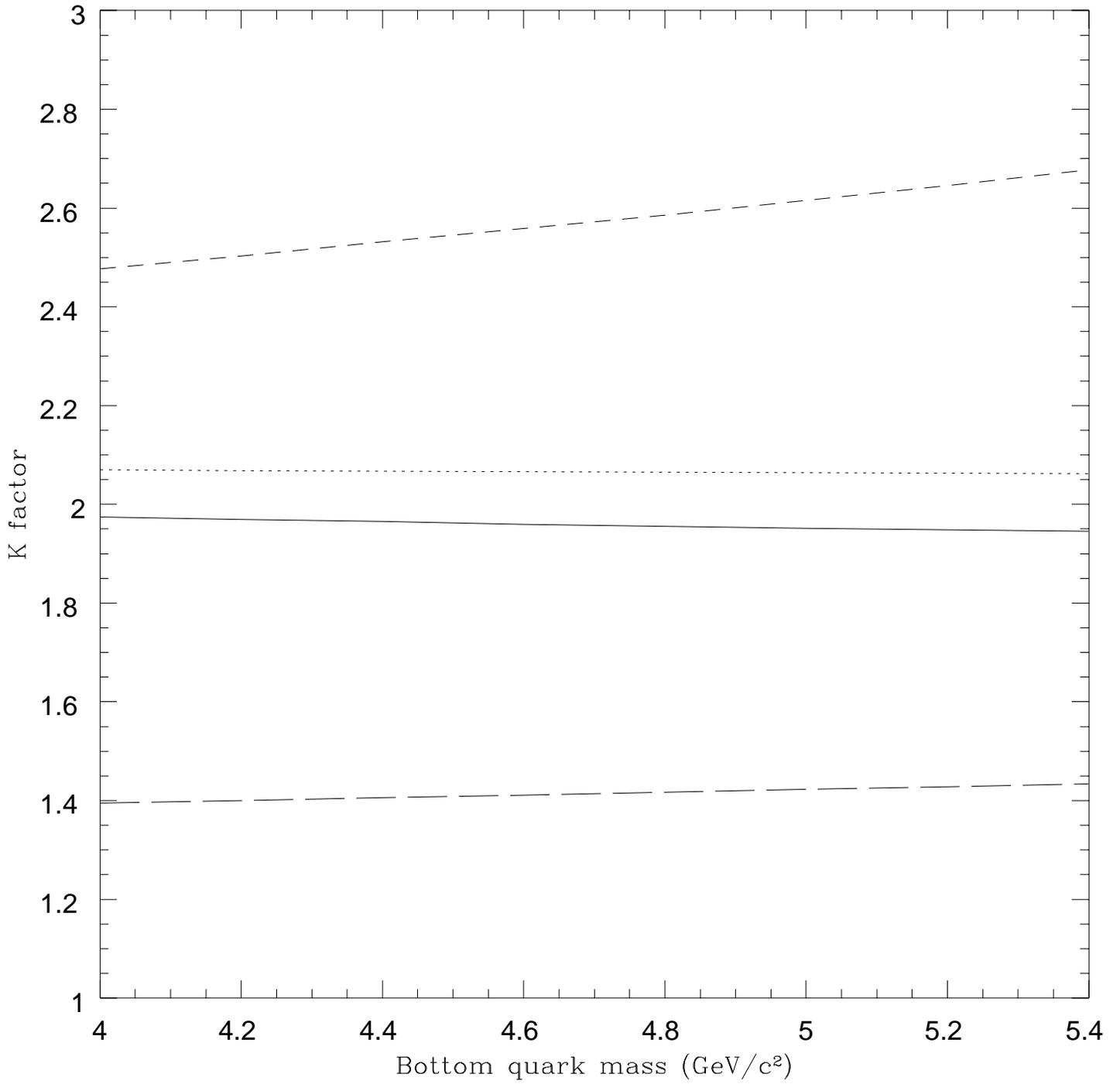
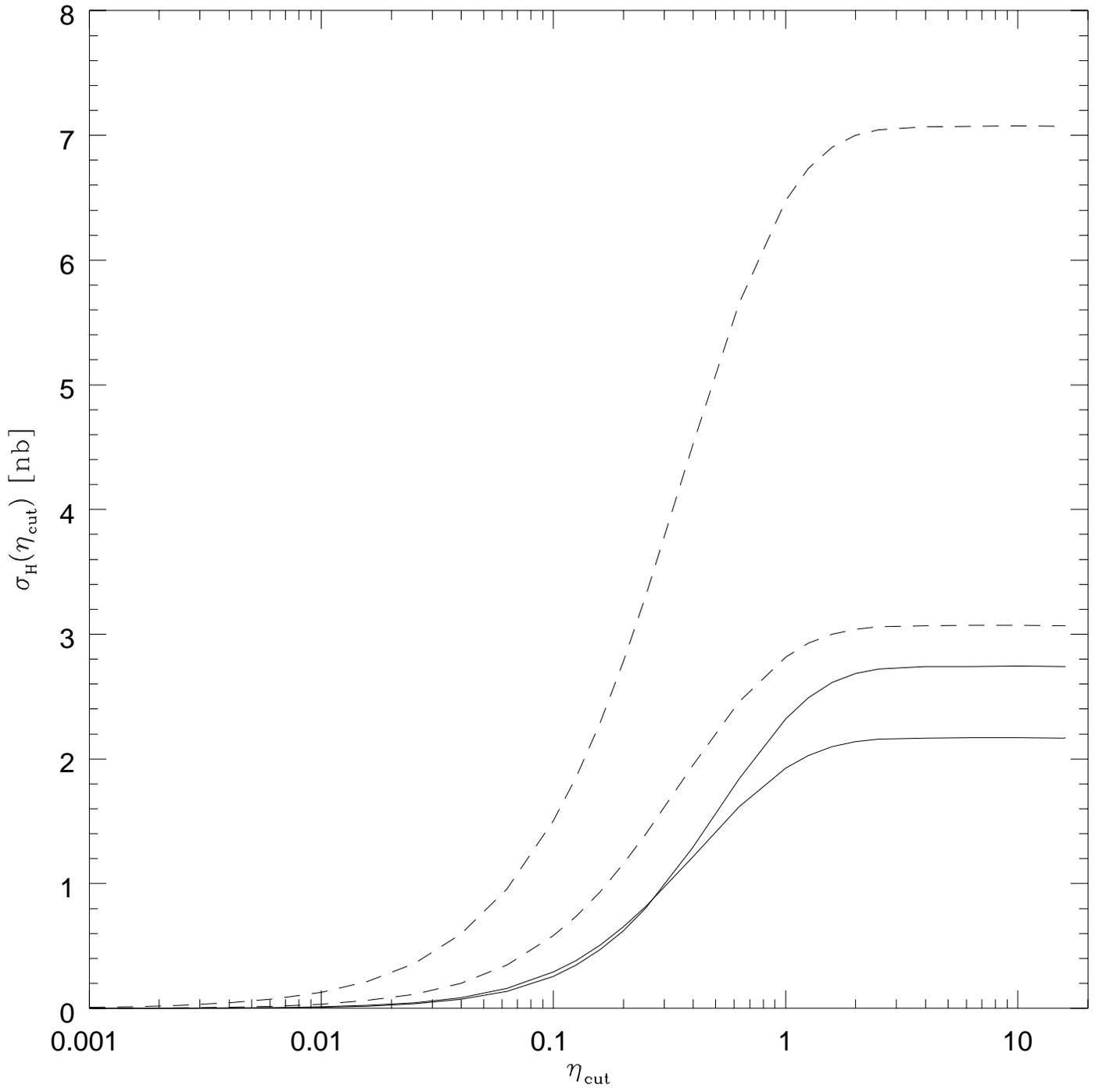
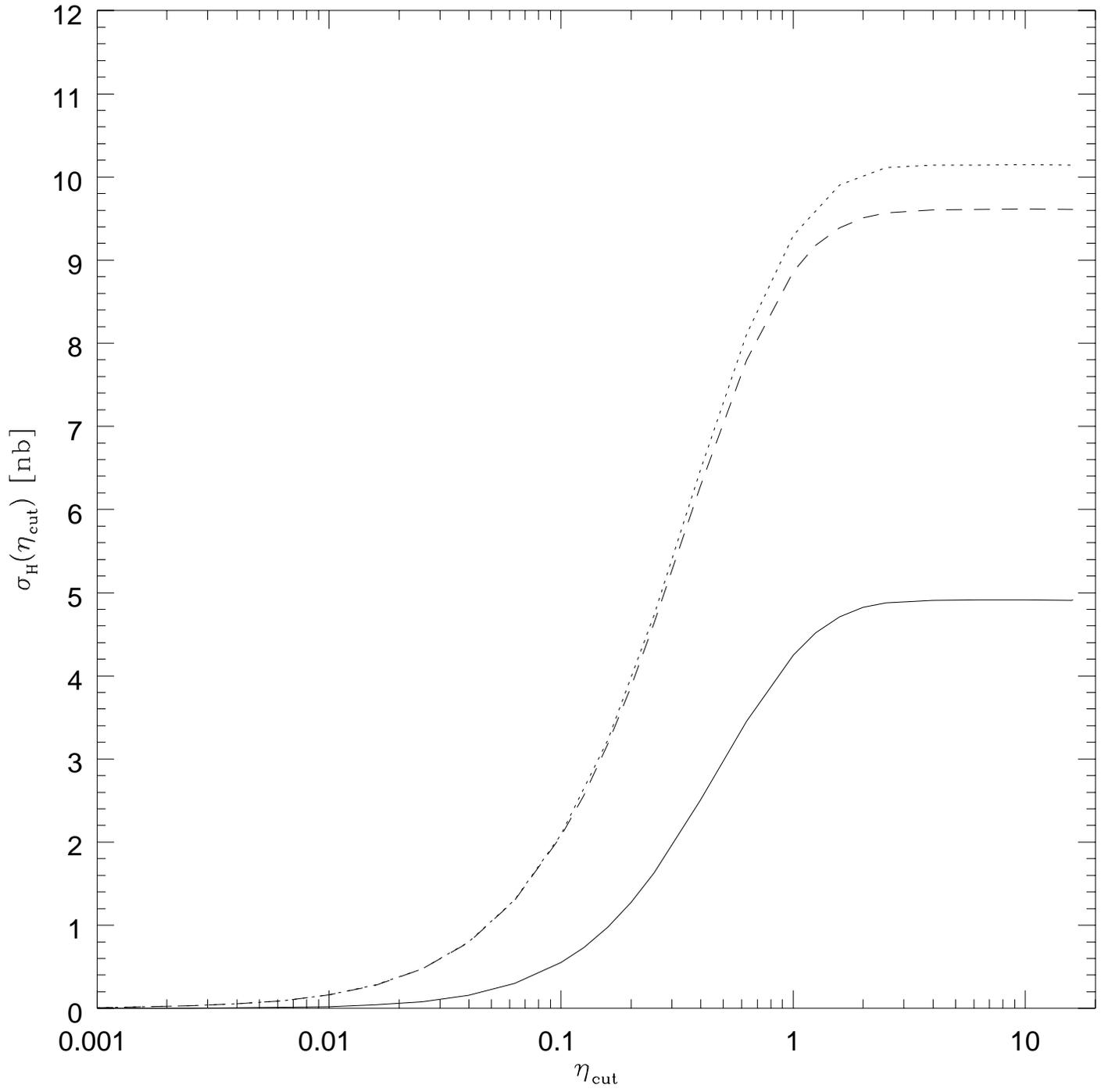
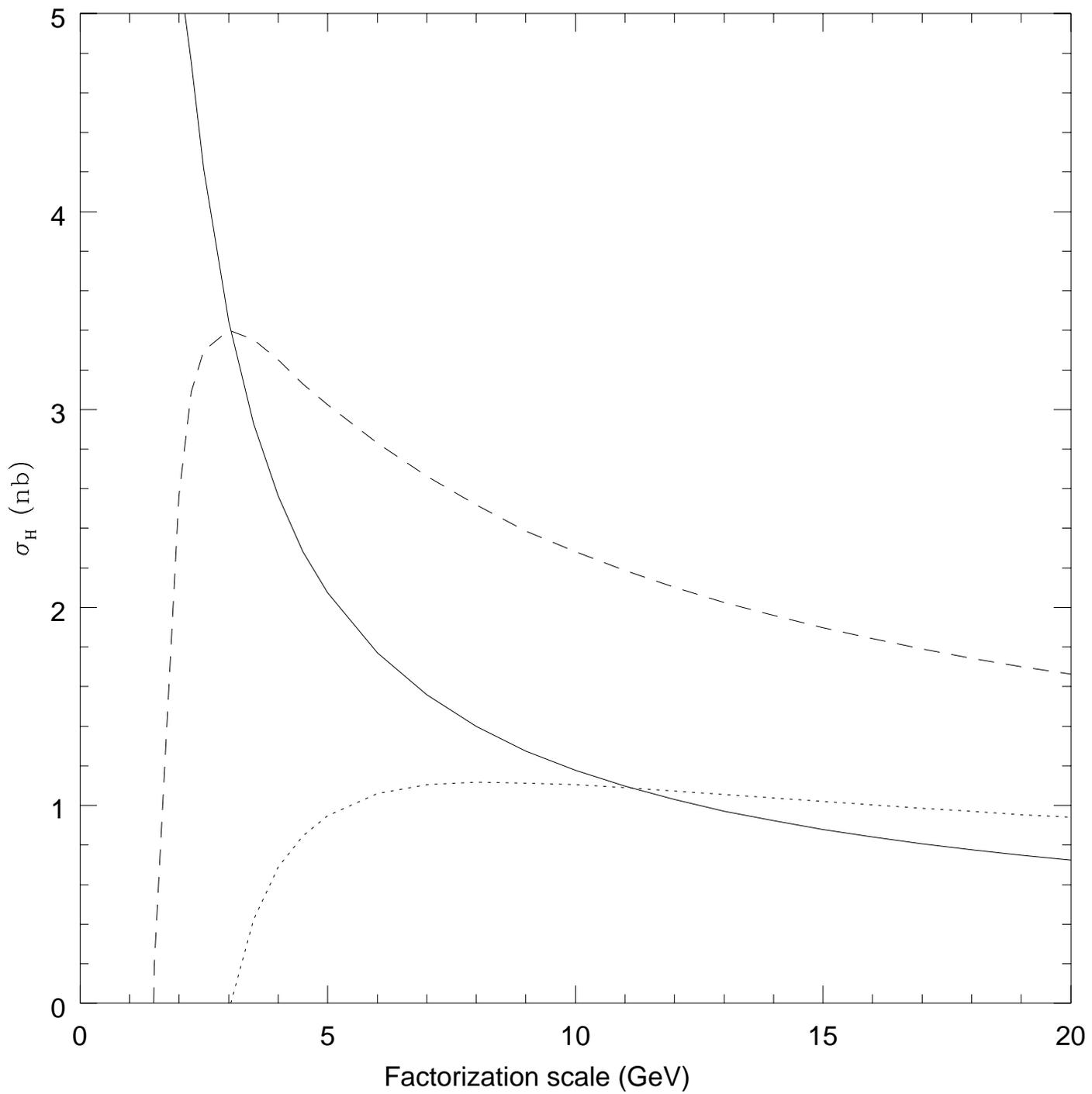
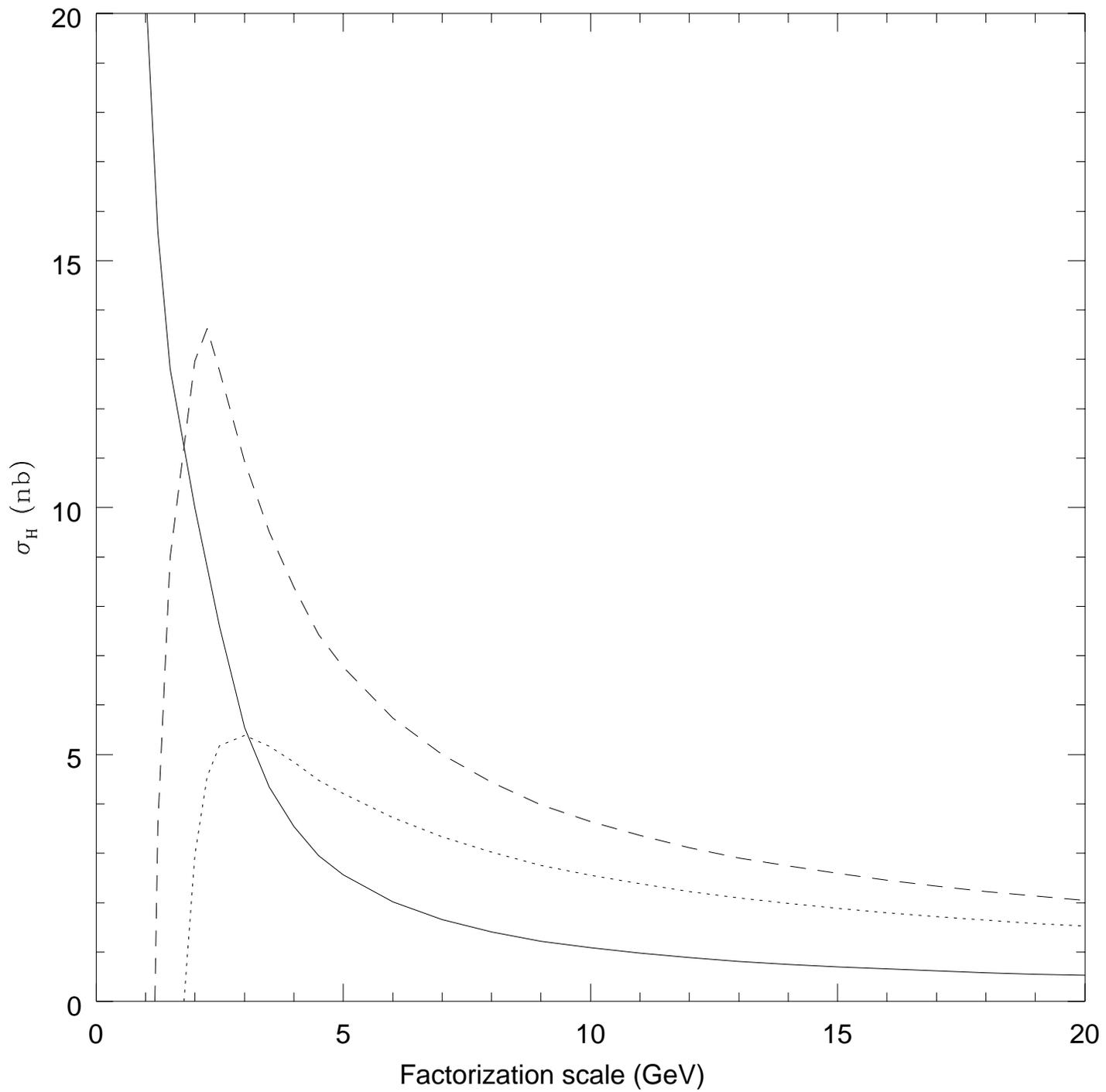
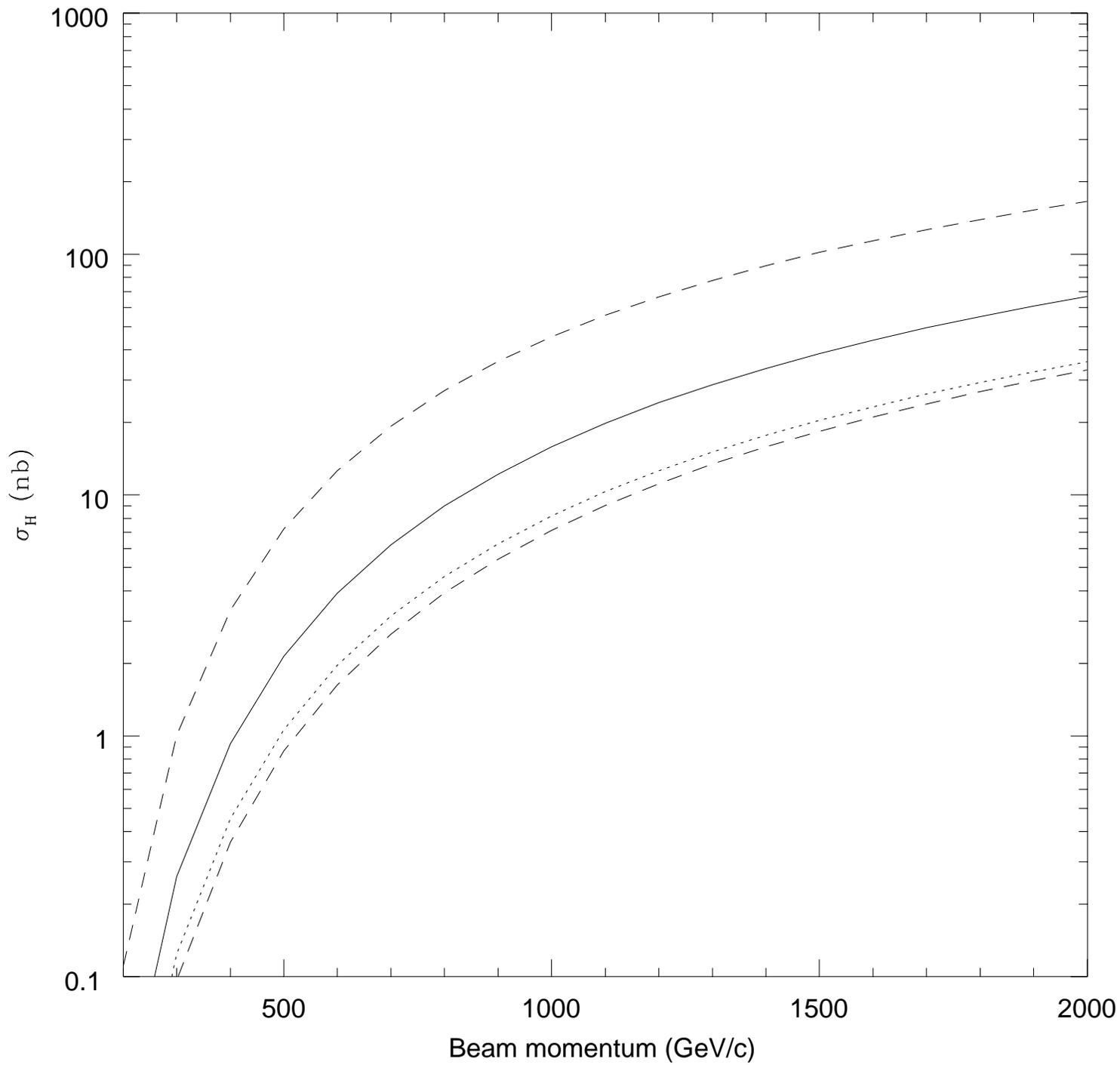
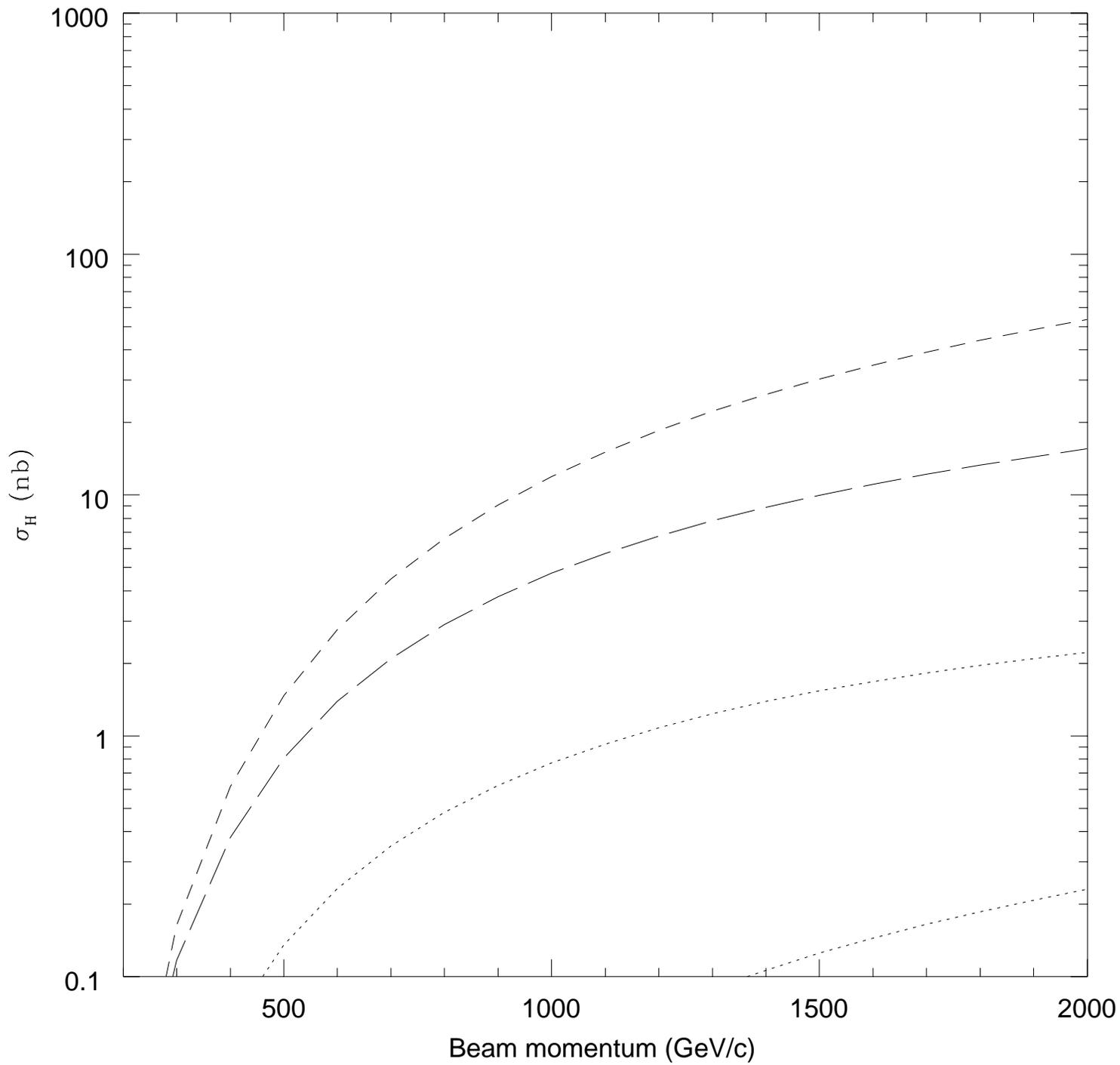
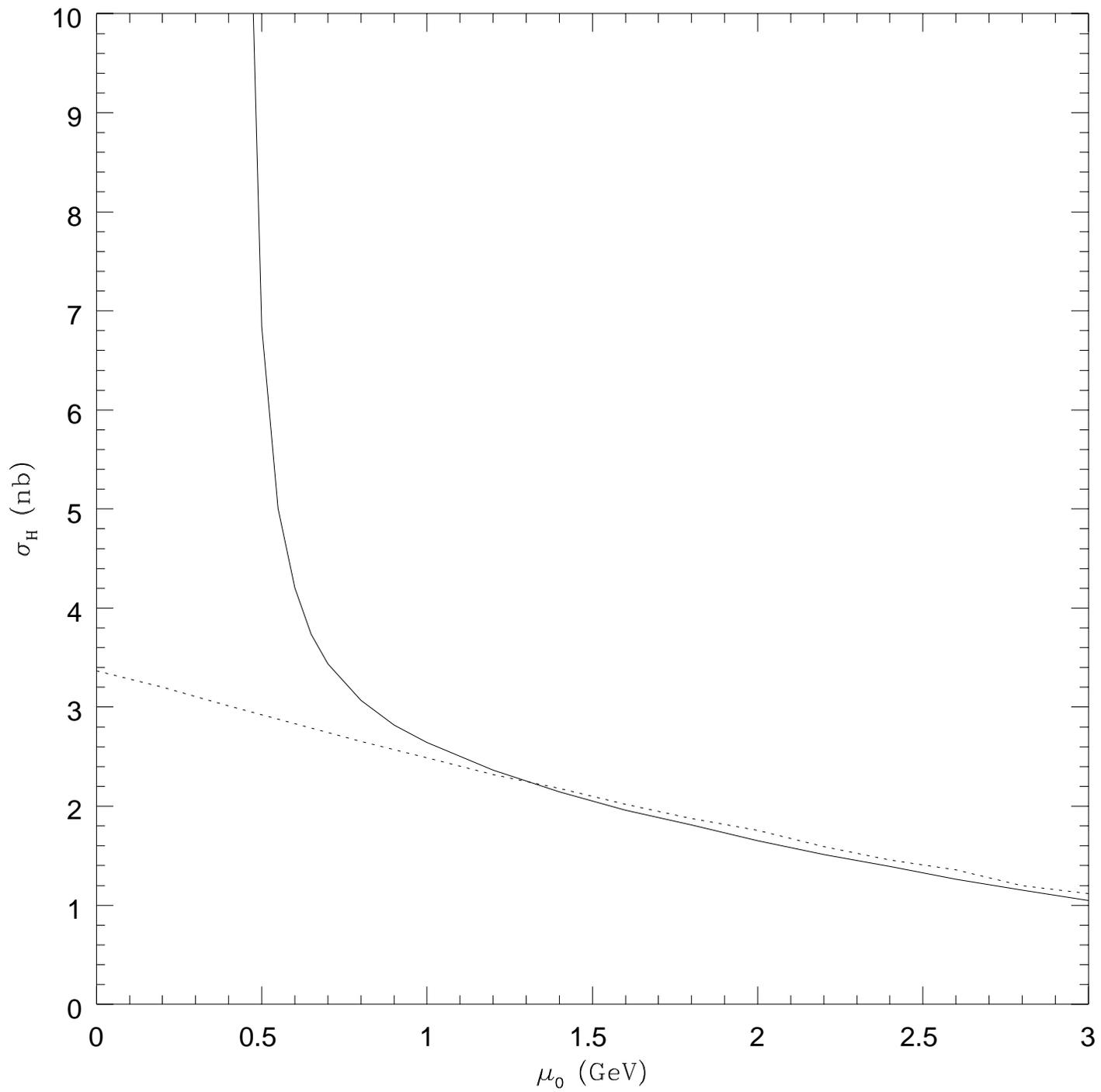
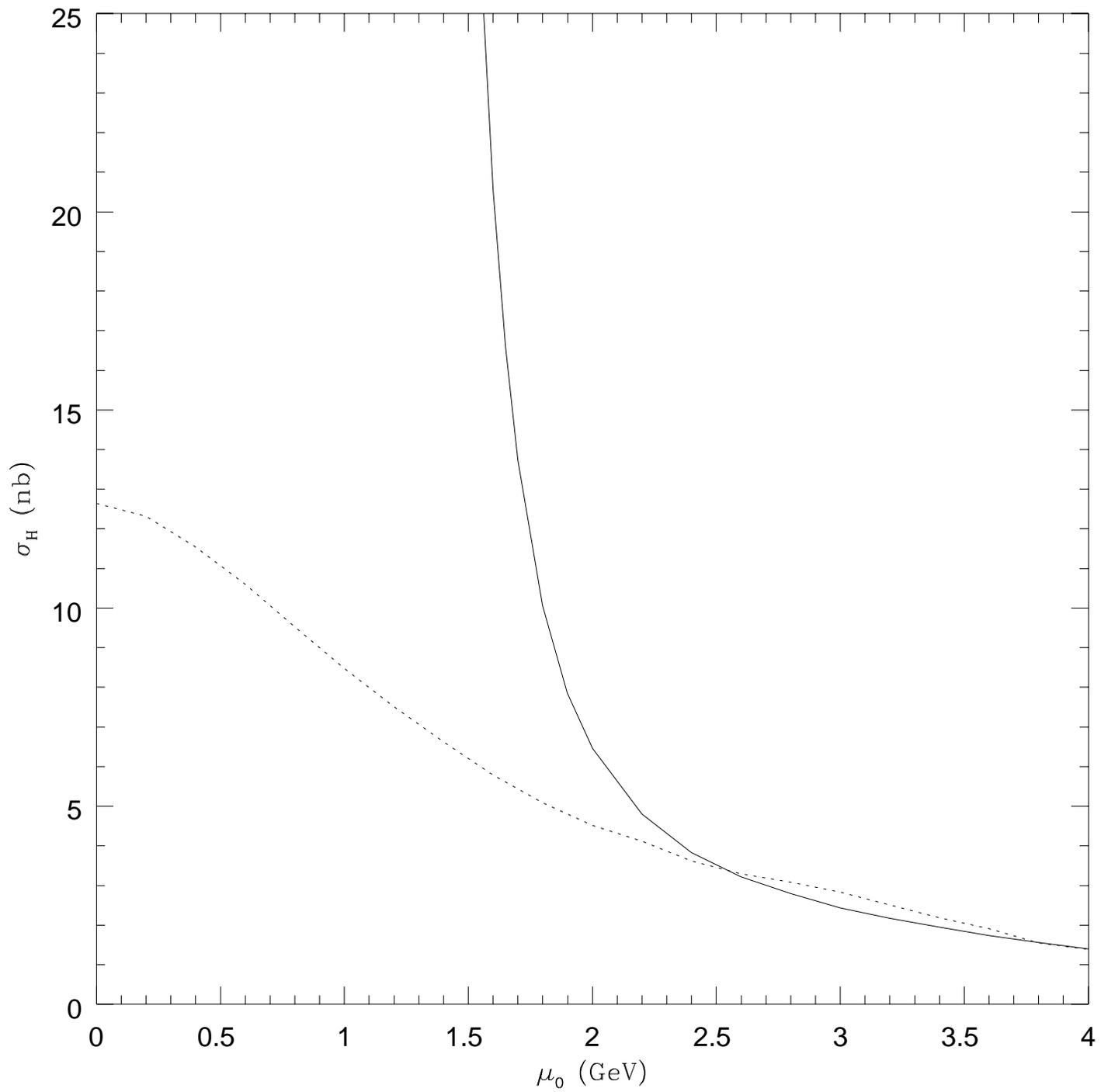
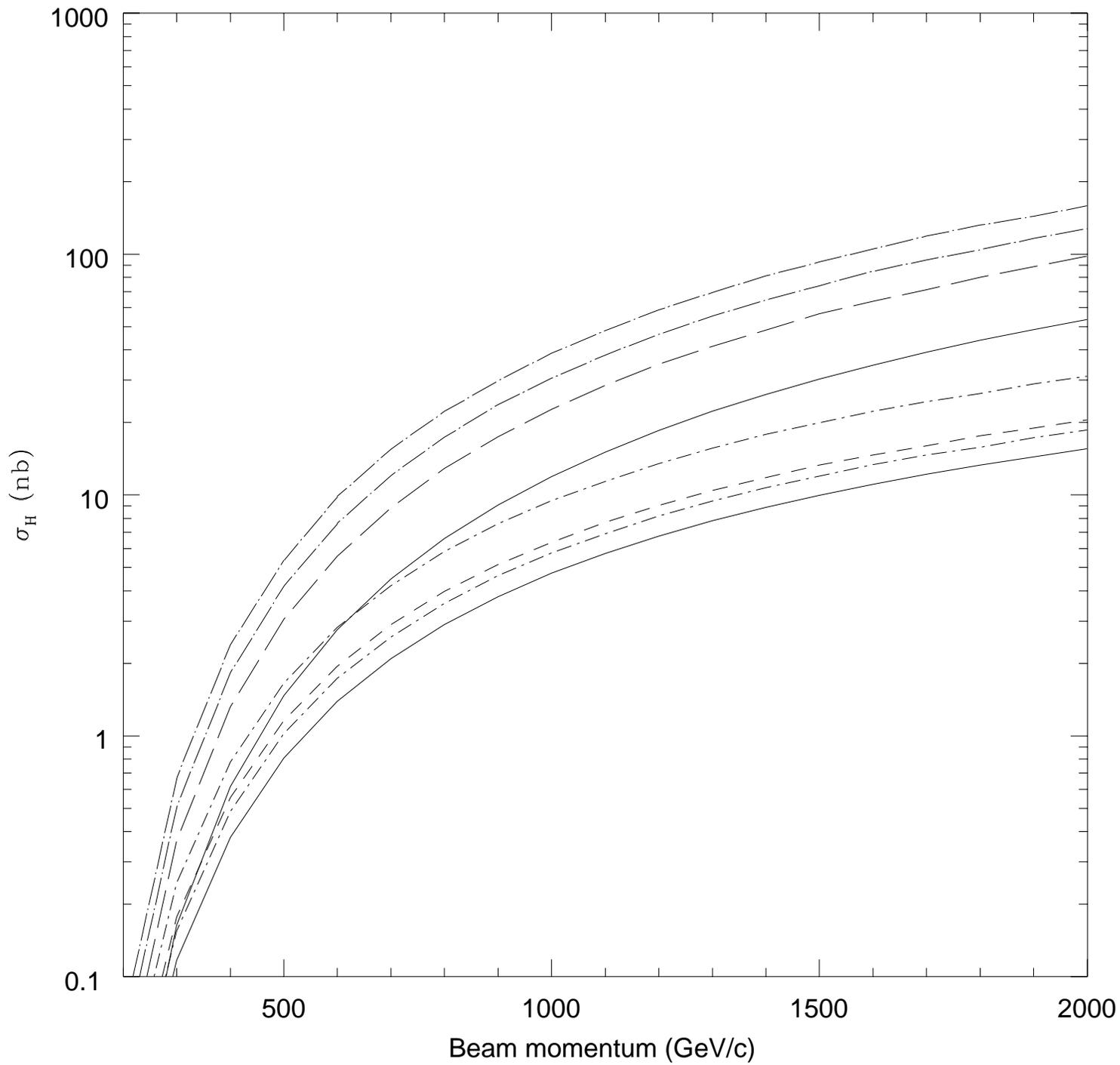
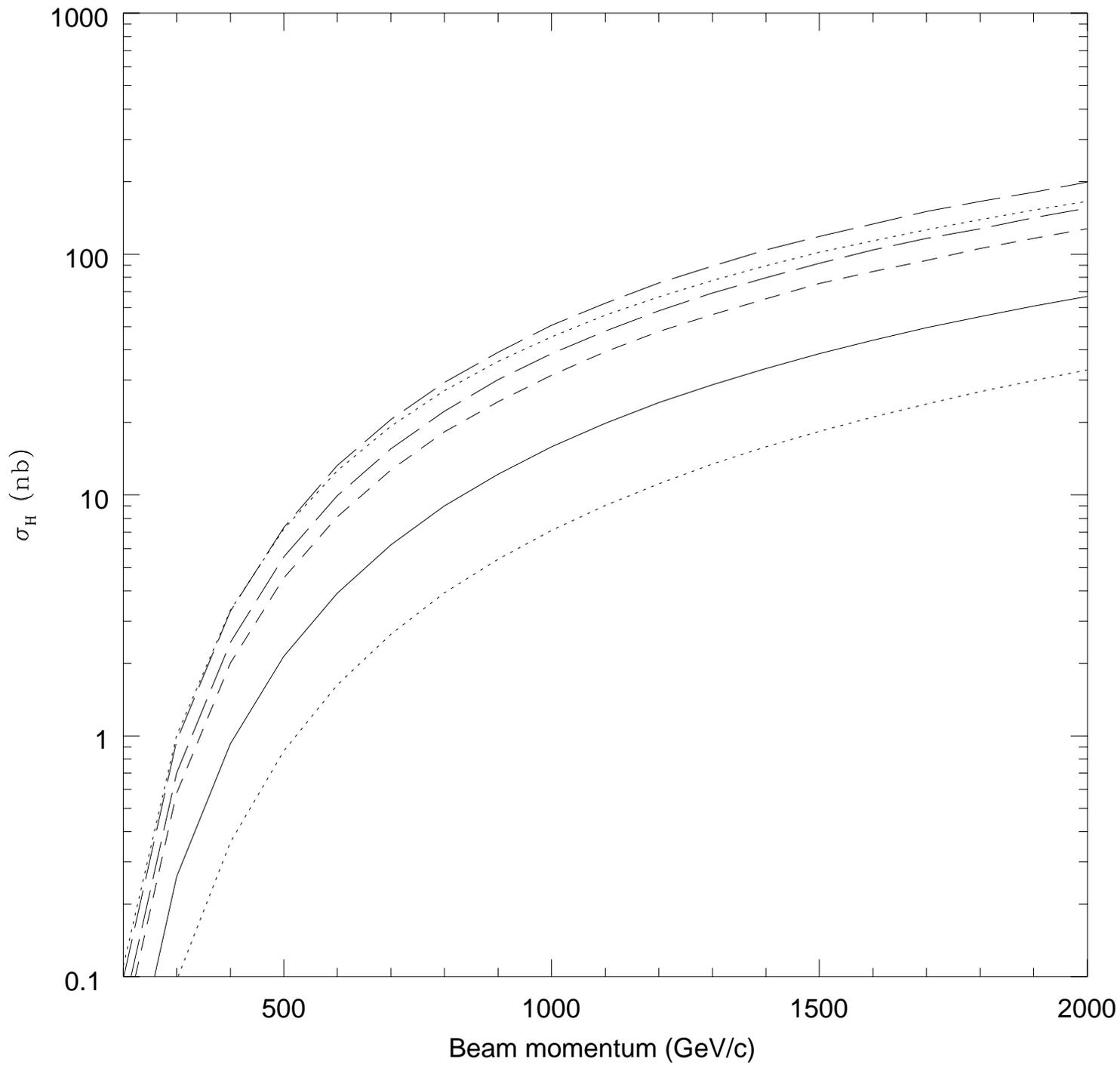
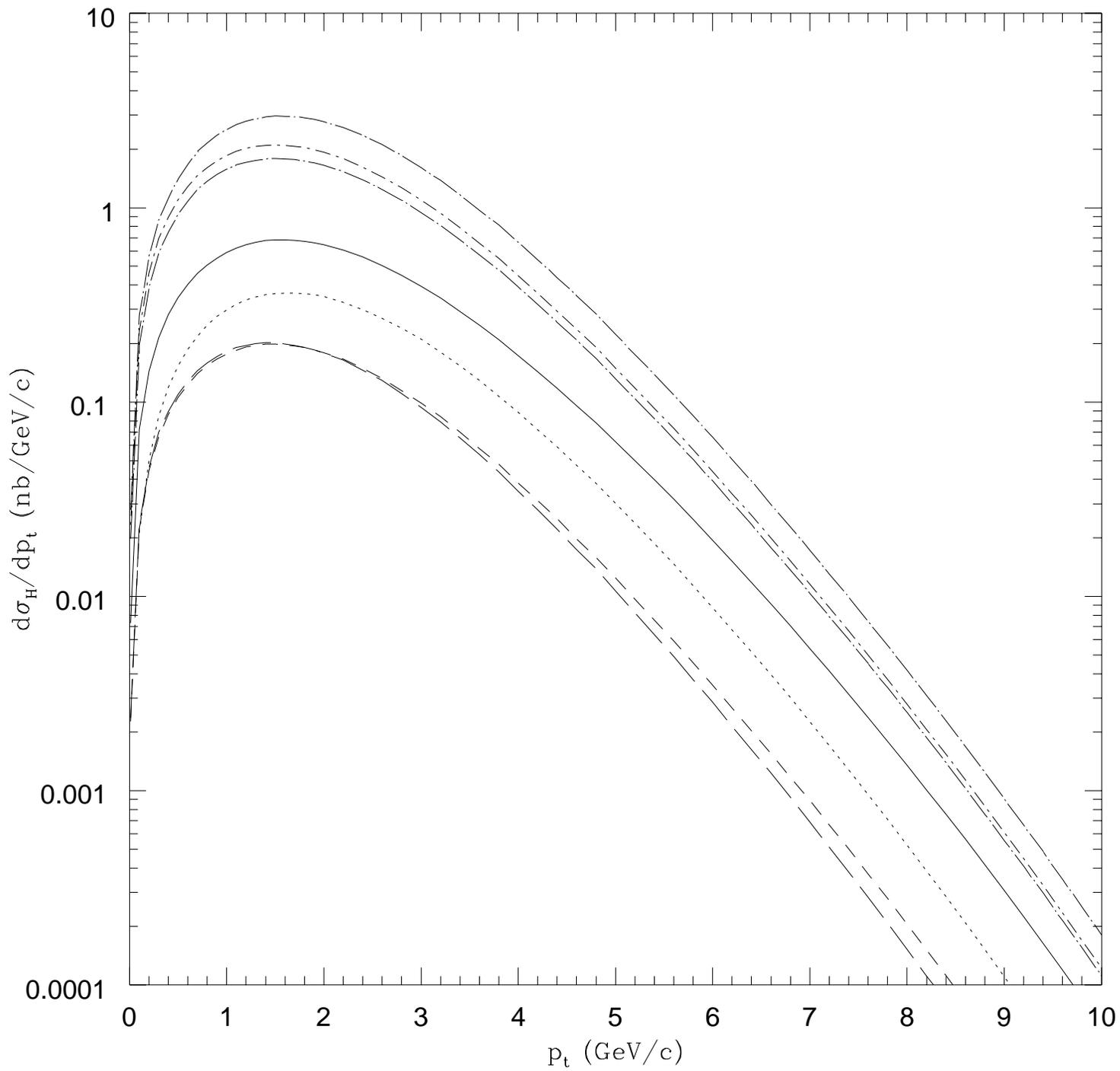
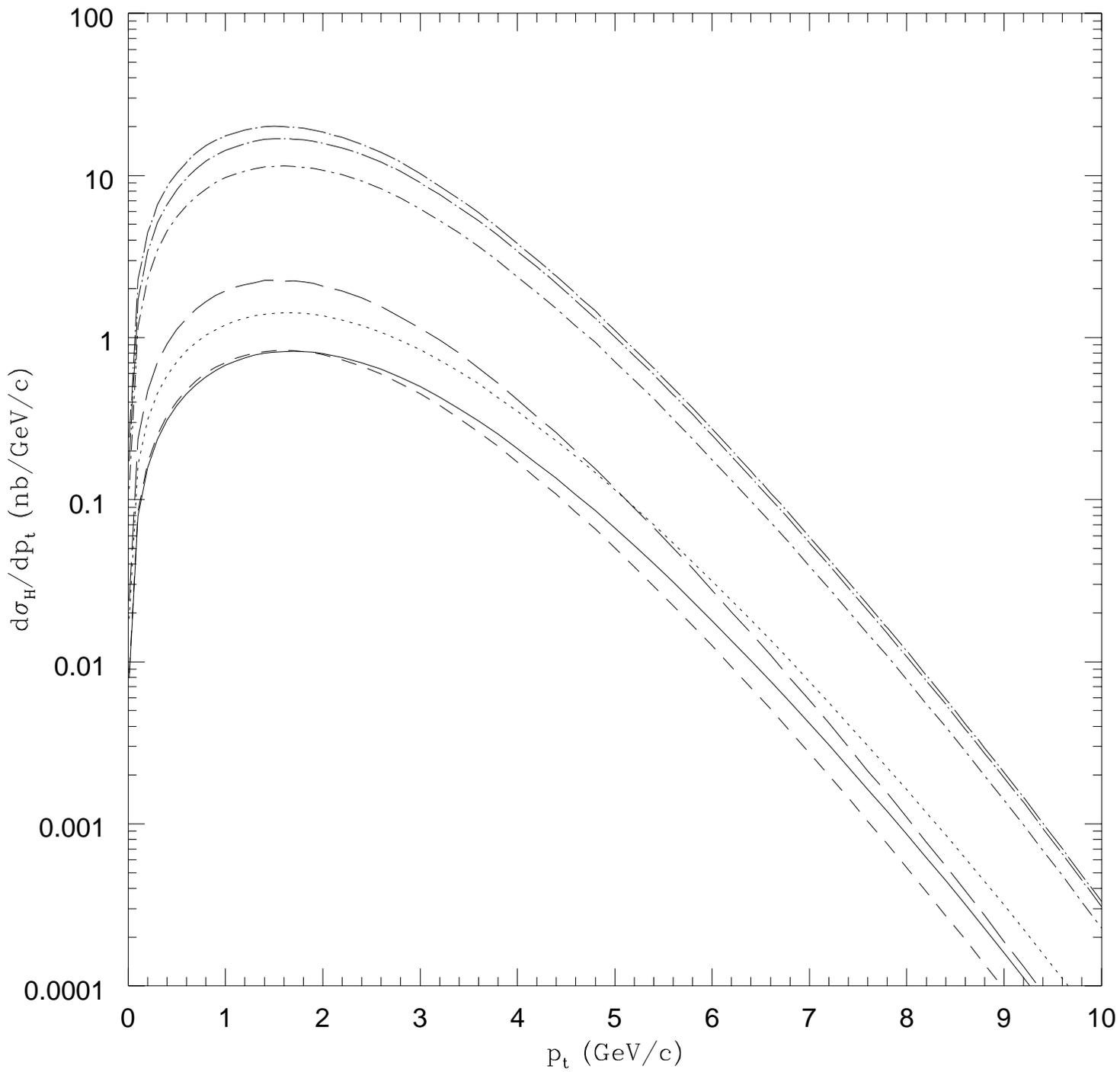
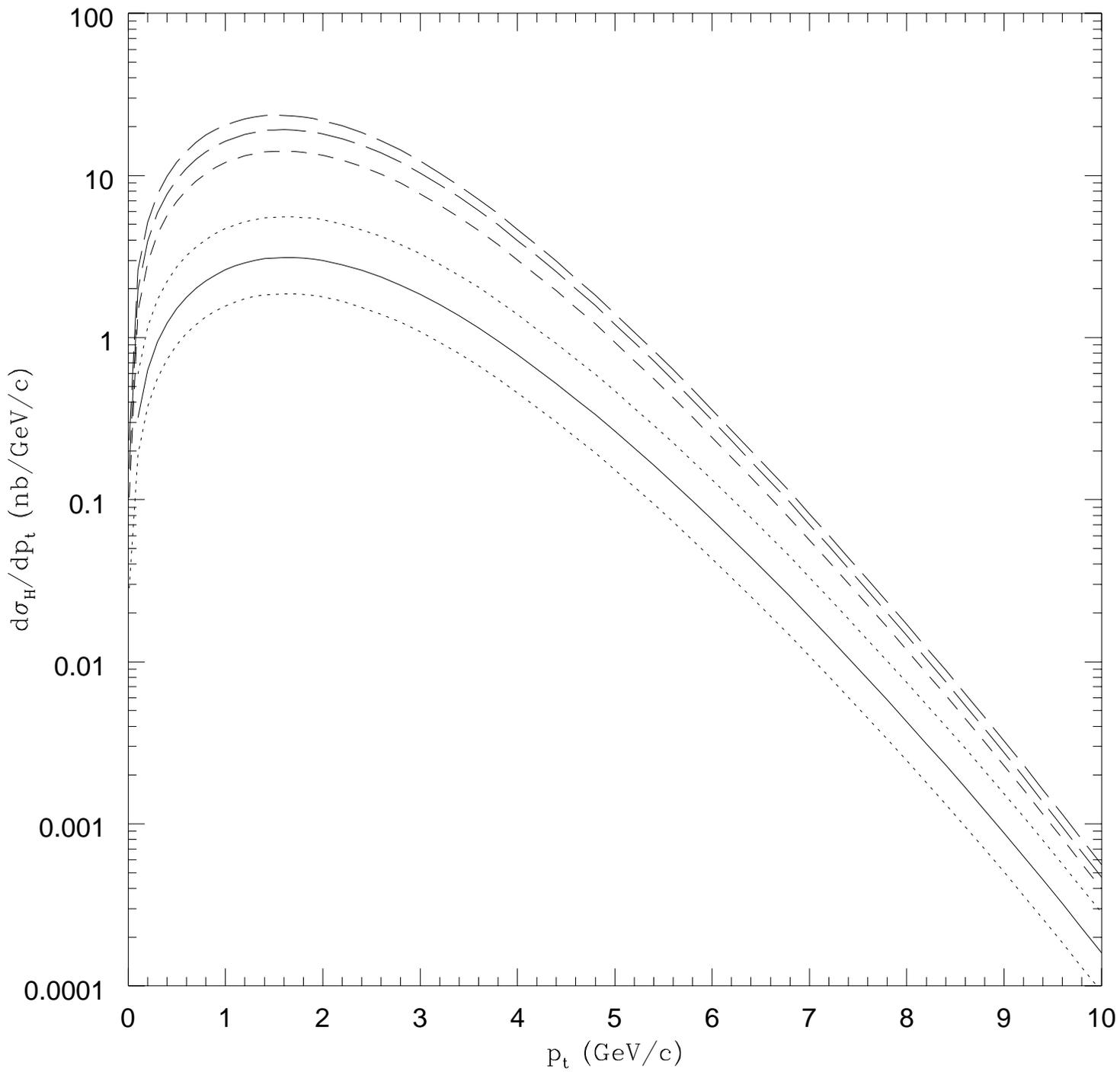
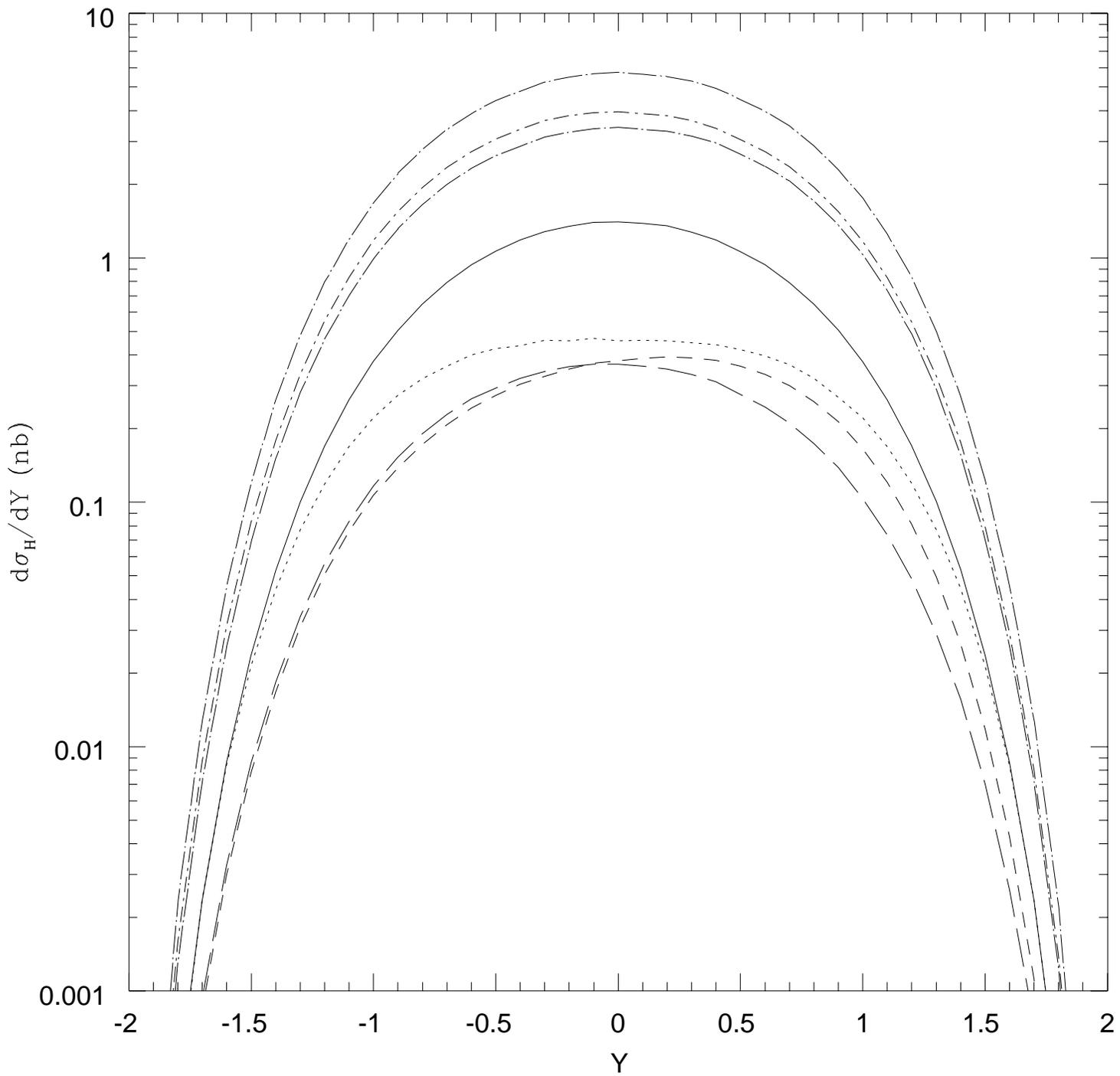
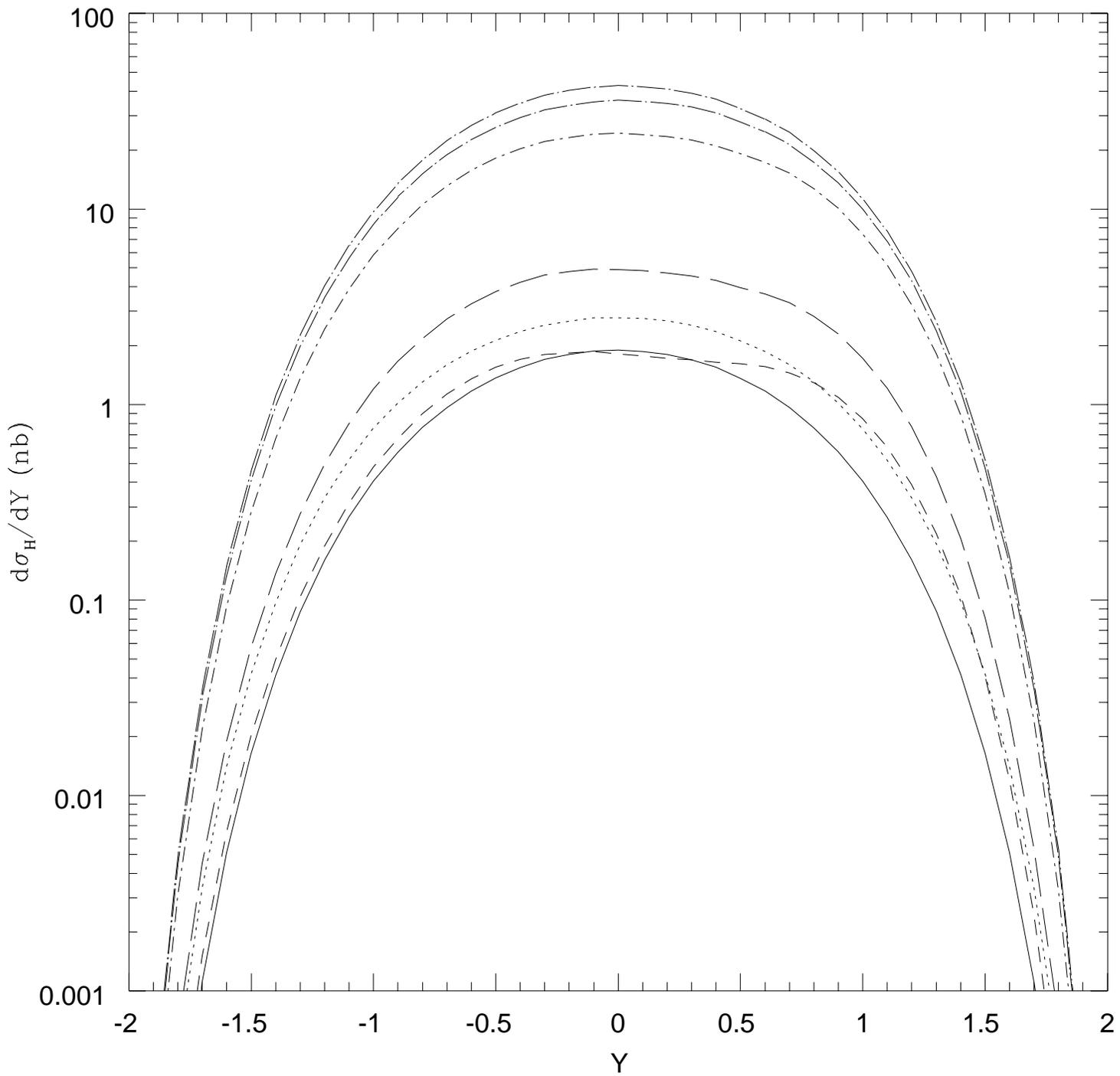
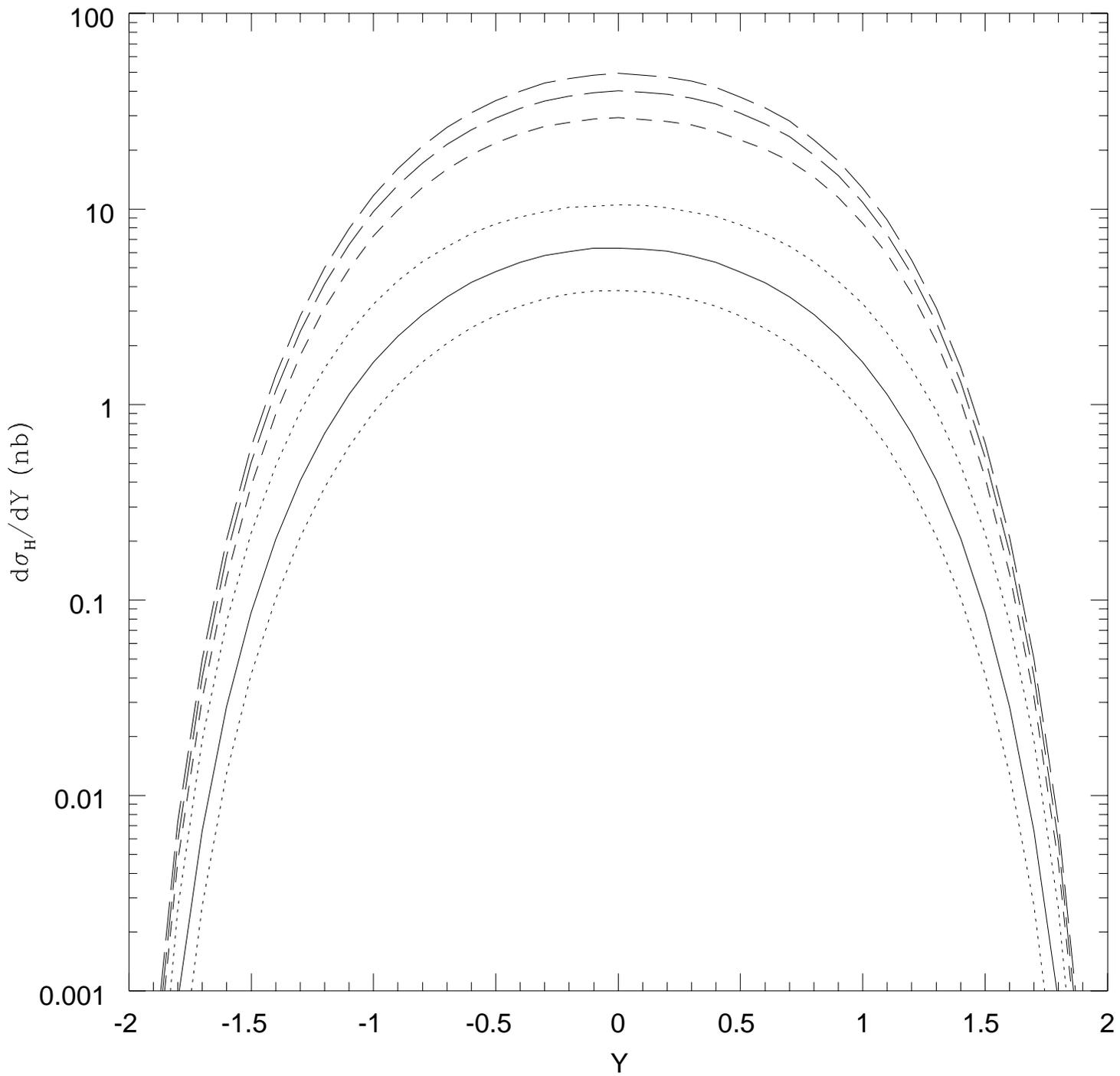
\end{document}